\documentclass[journal]{IEEEtran}

\ifCLASSINFOpdf
\usepackage{amsthm}
\usepackage{times}
\usepackage{epsfig}
\usepackage{graphicx}
\usepackage{amsmath}
\usepackage{amssymb}
\usepackage{comment}
\usepackage[font=small]{caption}
\usepackage{amsmath,amssymb} 
\usepackage{color,soul}
\usepackage{float}
\usepackage{booktabs}
\usepackage{multirow}
	\usepackage{textcomp}
\usepackage{array}
\graphicspath{{./figures/}}
\usepackage[comma,sort&compress,numbers]{natbib}
\usepackage[font=small]{caption}
\newcolumntype{P}[1]{>{\centering\arraybackslash}p{#1}}
\newcolumntype{M}[1]{>{\centering\arraybackslash}m{#1}}
\usepackage{url}
\usepackage{siunitx}
\usepackage{scrextend}
   \graphicspath{{./figures/}}
\else
\fi
\allowdisplaybreaks

\begin{document}
\title{CVEGAN: A Perceptually-inspired GAN for Compressed Video Enhancement}

\author{Di~Ma, Fan~Zhang and~David~R.~Bull
\thanks{Di Ma, Fan Zhang, and David R. Bull are with the Bristol Vision Institute, University of Bristol, Bristol, UK. }
\thanks{E-mail: \{di.ma, fan.zhang, dave.bull\}@bristol.ac.uk}}
\maketitle

\begin{abstract}

We propose a new Generative Adversarial Network for Compressed Video quality Enhancement (CVEGAN). The CVEGAN generator benefits from the use of a novel Mul\textsuperscript{2}Res block (with multiple levels of residual learning branches), an enhanced residual non-local block (ERNB) and an enhanced convolutional block attention module (ECBAM). The ERNB has also been employed in the discriminator to improve the representational capability. The training strategy has also been re-designed specifically for video compression applications, to employ a relativistic sphere GAN (ReSphereGAN) training methodology together with new perceptual loss functions. The proposed network has been fully evaluated in the context of  two typical video compression enhancement tools: post-processing (PP) and spatial resolution adaptation (SRA). CVEGAN has been fully integrated into the MPEG HEVC video coding test model (HM16.20) and experimental results demonstrate significant coding gains (up to 28\% for PP and 38\% for SRA compared to the anchor)  over existing state-of-the-art architectures for both coding tools across multiple datasets. 

\end{abstract}


\IEEEpeerreviewmaketitle

\section{Introduction}

Over the past decade, machine learning methods based on deep neural networks have provided revolutionary advances across various computer vision applications, in particular for image/video processing and understanding. More recently these have been successfully applied to picture compression, both in terms of enhancing individual tools within conventional codecs, and also in providing new end-to-end compression via auto-encoder architectures \cite{bull2021intelligent}. 

In the context of learning-based video coding tools, there is a distinct class of methods which provide superior performance by employing Convolutional Neural Network (CNN) processes to enhance the quality of the video reconstructed at the decoder. Typical examples include post-processing (PP) and video format adaptations (with CNN format restoration) \cite{ma2019image,liu2020deep}. These approaches often employ relatively simple network structures and utilise pixel-wise loss functions, which do not reflect the latest advances in the field. More recently,  compression enhancement tools have further improved perceptual quality using Generative Adversarial Networks (GANs) \cite{ma2019perceptually,wang2020multi}. In this case, the loss functions have typically combined pixel-wise distortions, simple quality metrics and feature map differences with artificially configured weights, which do not offer optimal correlation with visual quality.

\begin{figure*}[ht]
\centering
\scriptsize
\centering
\begin{minipage}[b]{0.16\linewidth}
\centering
\centerline{\includegraphics[width=1.01\linewidth]{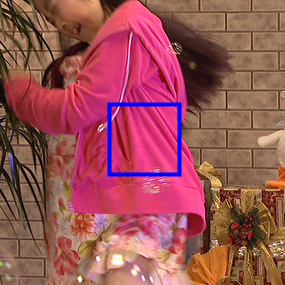}}
(a) Original 
\end{minipage}
\begin{minipage}[b]{0.16\linewidth}
\centering
\centerline{\includegraphics[width=1.01\linewidth]{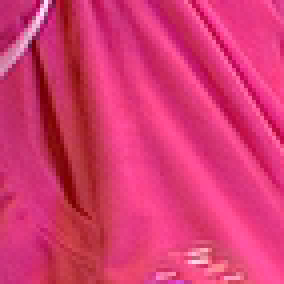}}
(b) Original
\end{minipage}
\begin{minipage}[b]{0.16\linewidth}
\centering
\centerline{\includegraphics[width=1.01\linewidth]{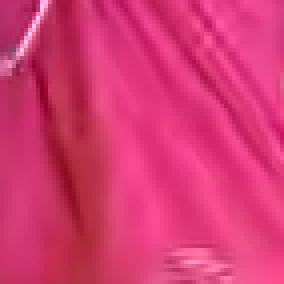}}
(c) HM 16.20, QP=37
\end{minipage}
\begin{minipage}[b]{0.16\linewidth}
\centering
\centerline{\includegraphics[width=1.01\linewidth]{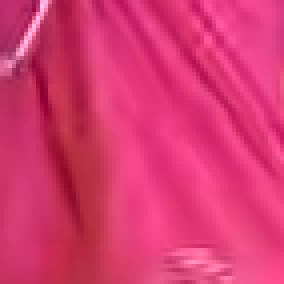}}
(d) RNAN \cite{zhang2019nonlocal}
\end{minipage}
\begin{minipage}[b]{0.16\linewidth}
\centering
\centerline{\includegraphics[width=1.01\linewidth]{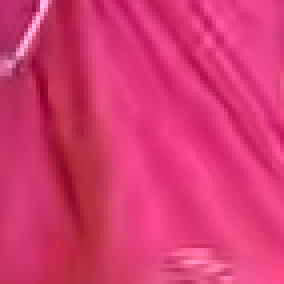}}
(e) MSRGAN \cite{ma2019perceptually}
\end{minipage}
\begin{minipage}[b]{0.16\linewidth}
\centering
\centerline{\includegraphics[width=1.01\linewidth]{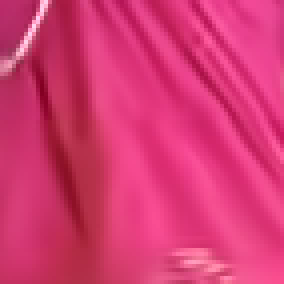}}
(f) CVEGAN (Ours)
\end{minipage}

\begin{minipage}[b]{0.16\linewidth}
\centering
\centerline{\includegraphics[width=1.01\linewidth]{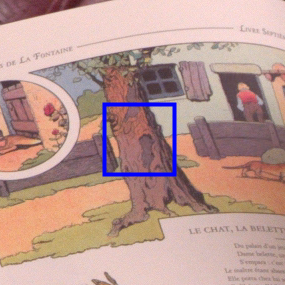}}
(g) Original
\end{minipage}
\begin{minipage}[b]{0.16\linewidth}
\centering
\centerline{\includegraphics[width=1.01\linewidth]{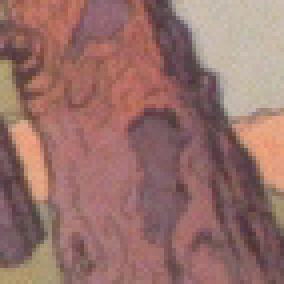}}
(h) Original
\end{minipage}
\begin{minipage}[b]{0.16\linewidth}
\centering
\centerline{\includegraphics[width=1.01\linewidth]{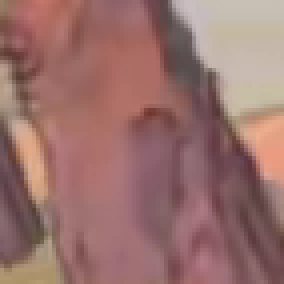}}
(i) HM 16.20, QP=37 
\end{minipage}
\begin{minipage}[b]{0.16\linewidth}
\centering
\centerline{\includegraphics[width=1.01\linewidth]{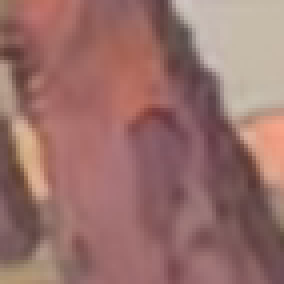}}
(j) RNAN \cite{zhang2019nonlocal}
\end{minipage}
\begin{minipage}[b]{0.16\linewidth}
\centering
\centerline{\includegraphics[width=1.01\linewidth]{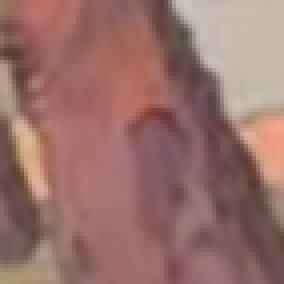}}
(k) MSRGAN \cite{ma2019perceptually}
\end{minipage}
\begin{minipage}[b]{0.16\linewidth}
\centering
\centerline{\includegraphics[width=1.01\linewidth]{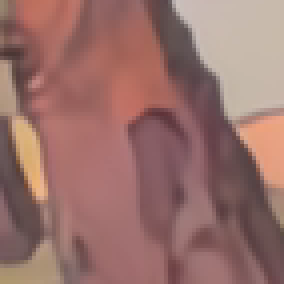}}
(l) CVEGAN (Ours)
\end{minipage}

\caption{Two sets of example blocks cropped from the reconstructed frames generated by the anchor HM 16.20 (QP=37), RNAN {\cite{zhang2019nonlocal}}, MSRGAN {\cite{ma2019perceptually}} and the proposed CVEGAN. The bitstreams for each example set consumed identical or similar bit rates. Row 1 corresponds to the 250th frame of the \textit{PartyScene} sequence (for CNN-based PP) and Row 2 corresponds to the 104th frame of the \textit{CatRobot1}  sequence (for CNN-based SRA). It can be observed that the output of CVEGAN exhibits improved perceptual quality compared to the anchor HEVC HM 16.20, RNAN {\cite{zhang2019nonlocal}} and MSRGAN {\cite{ma2019perceptually}}, with fewer blocking artefacts, more textural detail and higher contrast.}

\label{fig:perceptualcom}
\end{figure*}

In this context, we propose a novel GAN architecture for Compressed Video Enhancement (CVEGAN). The main contributions of this work are summarised below.

     (1) We present a novel Mul\textsuperscript{2}Res block structure to improve the network \textit{cardinality}, which employs multiple levels of multiple residual learning branches. The residual blocks in each branch also contain convolutional layers with various kernel sizes. This architecture improves overall enhancement performance compared to conventional residual learning blocks with a fixed kernel size. 
    
    (2) We employ enhanced residual non-local blocks (ERNBs) and enhanced convectional block attention modules (ECBAMs), both of which have been modified compared to their original structures \cite{zhang2019nonlocal,woo2018cbam}. Through extracting non-local features (ERNB) and applying channel and spatial attention mechanisms (ECBAM), the representational capability of the network has been further improved.
    
    (3) We have designed a new training methodology, Relativistic SphereGAN (ReSphereGAN), which embodies an enhanced version of the original SphereGAN \cite{park2019sphere}. ReSphereGAN considers both north pole-centred geodesic distance and the relativistic geodesic distance between the real and fake feature points in the hypersphere. This  stabilises the training process and further minimises the difference between generator output and training target. 
    
    (4) We also propose a novel perceptual loss function which optimises video quality during training. This linearly combines the logarithms of four existing losses, where the combining weights have been determined using data from eight subjective video quality databases.  
    
    (5) We conducted a comprehensive comparison between CVEGAN and several state-of-the-art architectures. This was done for two typical coding enhancement tools: post-processing and spatial resolution adaptation in the context of the MPEG HEVC coding standard. Results demonstrate that CVEGAN provides superior coding performance compared to its counterparts based on both objective quality assessments and subjective comparisons. Figure \ref{fig:perceptualcom} shows a subjective comparison of example blocks cropped from the reconstructed frames produced by the anchor HEVC codec, CVEGAN and other state-of-the-art architectures.

\section{Related Work}

With dramatic increases in numbers of users and available content, the introduction of new immersive formats and higher user quality expectations, video content is now by far the greatest consumer of global internet bandwidth. These increased demands are challenging all delivery technologies, not least video compression which plays a key role in the trade-off between limited network capacity and elevated user video quality expectations. 

\subsection{Standard Video Codecs}

The latest video coding standard, Versatile Video Coding (VVC) \cite{bross2018versatile},  approved by ISO and ITU in 2020,  provides 30-40\% bitrate savings over its predecessor, the High Efficiency Video Coding (HEVC) \cite{hevc}. A competitor of HEVC and VVC is the open source and royalty-free codec, AV1,  which was released by the Alliance for Open Media (AOM) in 2018. AV1 also achieves significant coding gains compared to HEVC \cite{zhang2020comparing}. Despite their evident coding gains,  none of these codecs have widely exploited machine learning methods to optimise their architecture or tools. In contrast several learning-based approaches have been published in the open literature which achieve very promising results compared to these standards.

\subsection{Deep Learning-based Video Compression}
\label{subsec:dvc}

Existing learning-based picture coding algorithms can be classified into two primary categories. The first relates to end-to-end training and optimisation using auto-encoder type architectures  \cite{balle2016end,jiang2017end,minnen2018joint,agustsson2019generative,mentzer2019practical,rippel2019learned,lu2019dvc,djelouah2019neural,habibian2019video,lin2020mlvc,agustsson2020scale}. Although solutions in this category are not yet competitive with the latest standardised codecs, such as VVC and AV1, they demonstrates significant potential for the future. 

A second class contains algorithms (primarily based on CNNs) that are designed to enhance individual coding tools within a standard codec configuration. Such approaches have been used to optimise tools including:  intra prediction \cite{yeh2018hevc,li2018fully}, motion estimation \cite{zhao2018enhanced,zhao2019enhanced}, transforms \cite{puri2017cnn,jimbo2018deep}, quantisation \cite{alam2015perceptual}, entropy coding \cite{song2017neural,ma2018convolutional}, post-processing (PP) \cite{zhao2019cnn,lin2019partition}, in-loop filtering (ILF) \cite{jia2019content,ma2020mfrnet} and format adaptation \cite{lin2018convolutional,afonso2019video,ma2020gan}.  

\begin{figure*}[ht]
\centering
\includegraphics[width=1.01\linewidth]{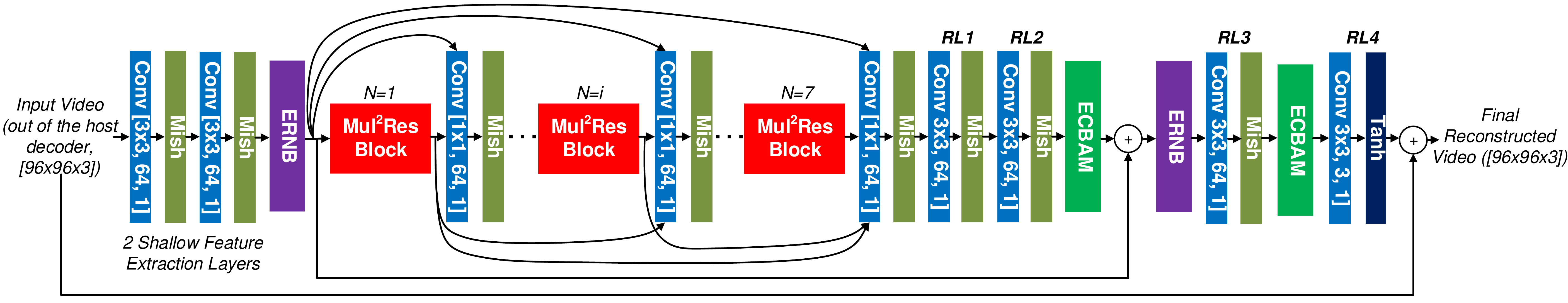}
\caption{Illustration of the CVEGAN's Generator (CVENet).}
\label{fig:cvenet}
\end{figure*}

Among these CNN-based coding tools, there is one group of methods, which stand out in offering higher coding gains compared to the others \cite{ma2019image,liu2020deep}. These perform CNN operations at the decoder to enhance the quality of reconstructed video frames. Typical examples include post-processing (PP) \cite{lin2019partition,zhang2020video}, in-loop filtering \cite{zhang2018residualInLoop,jia2019content}, spatial resolution adaptation (SRA) and bit depth adaptation \cite{zhang2019vistra2}. While most PP approaches focus on single frame enhancement, some recent contributions have reported multi-frame enhancement methods \cite{yang2018multi,wang2020multi}. However, because these are associated with much higher computational complexity, we address only single frame enhancement here.

\subsection{Network Architecture}
\label{subsec:reviewstructure}

Most CNN architectures reported for video enhancement have their origins in single image restoration. Their common architectural features include: (i) concatenated convolutional layers \cite{li2017cnn} (ii) concatenated residual blocks \cite{zhang2019vistra2}; (iii) residual dense connections \cite{ma2020gan}; (iv) cascading connections \cite{ma2020mfrnet}; and (v) feature review structures \cite{wang2019progressive}. More recently, new advanced structures have also been proposed including channel and spatial attention mechanisms \cite{woo2018cbam}  and non-local feature extraction \cite{wang2018non}.

The performance of a CNN is generally related to three primary factors: \textit{depth} (i.e. the depth of networks), \textit{width} (i.e. the number of feature maps), and \textit{cardinality} (i.e. the size of transformation sets) \cite{goodfellow2016deep,zagoruyko2016wide,xie2017aggregated}. Most of the network architectures described above have been designed to increase the network \textit{depth} \cite{kim2016accurate,ledig2017photo,wang2018esrgan,zhang2018residual,zhang2018image,zhang2019nonlocal,ma2019perceptually,shang2020rfbesrgan} and \textit{width} \cite{zagoruyko2016wide,huang2017densely,lim2017enhanced,zhang2018residual,wang2018esrgan,cai2020rcagan,shang2020rfbesrgan}, while only a few of them exploit the \textit{cardinality} characteristic \cite{szegedy2016rethinking,xie2017aggregated,chollet2017xception,zhang2020resnest}. However, \textit{cardinality} is widely acknowledged to be a more effective way to improve overall performance and network capacity compared to the other two factors \cite{xie2017aggregated}. We also note that the kernel size of convolutional layers in these existing networks is usually set at a fixed value (3 in most cases), which may limit the receptive field size and hence the overall network performance \cite{goodfellow2016deep}.

\subsection{Training Strategy}
\label{subsec:gan}

In most deep learning-based coding enhancement tools, $\ell 1$ or $\ell 2$ loss is used to train the CNN models with the aim of minimising pixel-wise distortions. Alternative training strategies have been proposed to improve perceptual image quality, typically based on Generative Adversarial Network (GAN) architectures and loss functions combining  $\ell 1$/$\ell 2$ loss, feature map differences (e.g. VGG19-54 \cite{wang2018esrgan}) and low-complexity quality metrics (e.g. MS-SSIM and SSIM). Notable examples include approaches using standard GANs \cite{ledig2017photo,lin2019adgan}, Relativistic average GANs (RaGAN) \cite{wang2018esrgan,ahn2019photo,ma2019perceptually,ma2020gan,zhang2020video,muhammad2020srrescgan,ren2020rcangan,shang2020rfbesrgan}, conditional GANs (cGAN) \cite{cai2020rcagan}, Patch GANs \cite{ji2020patchesrgan}, and Wasserstein GAN-gradient penalty (WGAN-GP) \cite{wang2019cfsnet}.

Moreover, it is well known that $\ell 1$ and $\ell 2$ losses do not correlate well with subjective video quality \cite{zhang2018bvi,zhang2018unreasonable,chen2020perceptually}, and the combined loss functions employed in these GAN-based training strategies use artificially configured combining weights, which have never been fully evaluated in terms of their correlation with subjective video quality. These issues inevitably lead to sub-optimal training performance when the networks are utilised for compression application.

\section{The Proposed Network Architectures}

The proposed CVEGAN follows the basic GAN framework \cite{goodfellow2014generative}, combining a generator and a discriminator. Its architecture is described in the following subsections.

\subsection{Generator Architecture}

The generator, denoted as CVENet, is shown in Figure \ref{fig:cvenet}. This takes a 96$\times$96 YCbCr 4:4:4 compressed image block as the input and outputs a processed image block in the same format, targeting its original uncompressed version. The kernel sizes, number of feature maps and stride values for each convolutional layer of CVENet are shown in Figure \ref{fig:cvenet}.

CVENet has three unique structural features: (1) Mul\textsuperscript{2}Res blocks; (2) enhanced residual non-local blocks (ERNB); (3) enhanced convolutional block attention modules (ECBAM). These are described below. It can be seen that a Mish activation function \cite{misra2019mish} has been employed after all convolutional operations except for the final one. Mish has been previously reported to offer better performance than other commonly used functions such as ReLU, leaky ReLU (LReLU) and parametric ReLU (PReLU) \cite{misra2019mish}. 

\begin{figure}[ht]
\centering
\includegraphics[width=0.85\linewidth]{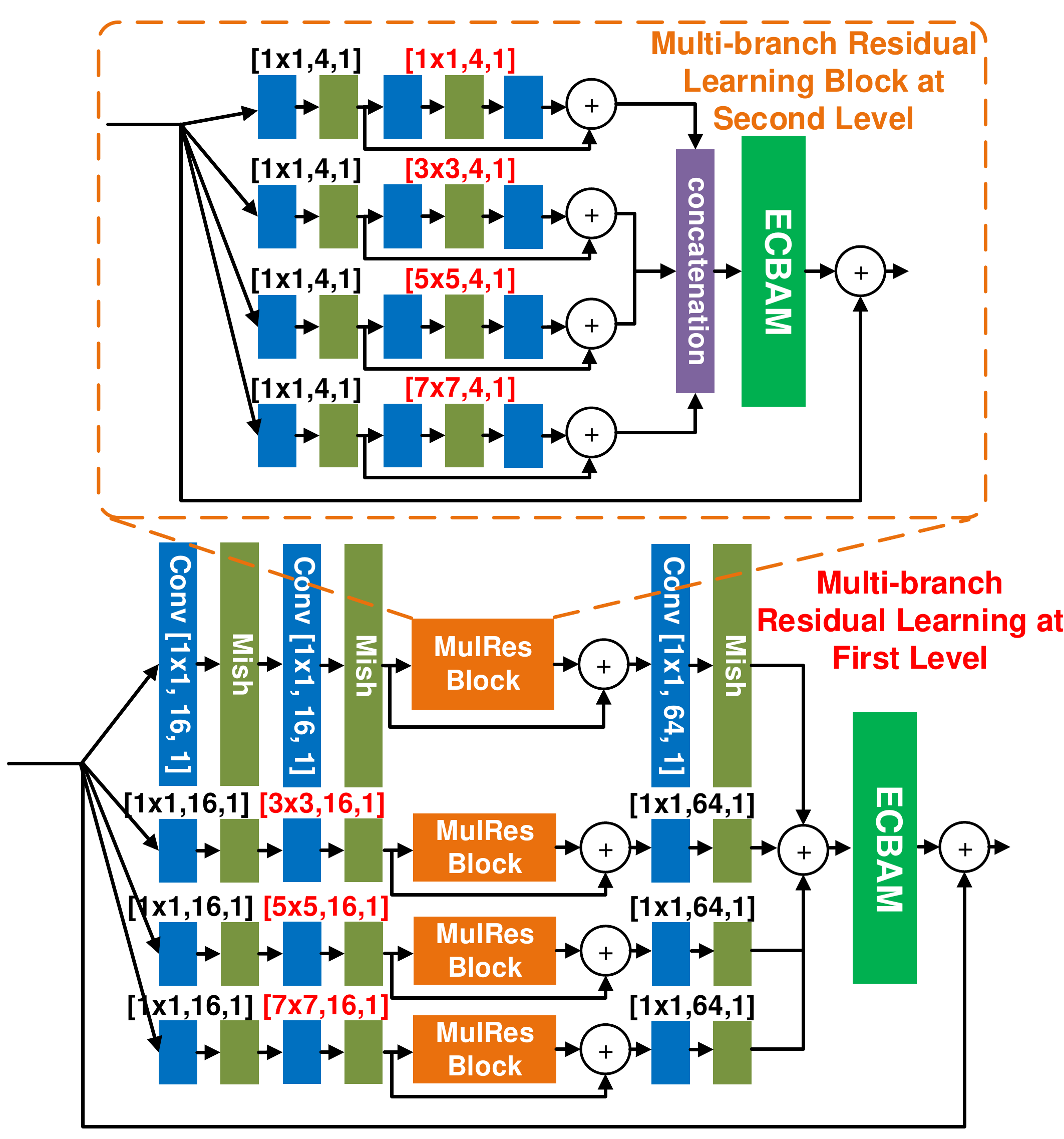}
\caption{Illustration of an Mul\textsuperscript{2}Res Block.}
\label{fig:mul2res2}
\end{figure}

In order to further improve information flow between Mul\textsuperscript{2}Res blocks, cascading connections \cite{ahn2018fast} (shown as black curves in Figure \ref{fig:cvenet}) are utilised between the input of the first Mul\textsuperscript{2}Res block and the 1$\times$1 convolutional layer after each of the Mul\textsuperscript{2}Res blocks, and between the output of each Mul\textsuperscript{2}Res block (except the final one) and the 1$\times$1 convolutional layer after each of the subsequent Mul\textsuperscript{2}Res blocks.

\textbf{Mul\textsuperscript{2}Res Block}: The new Mul\textsuperscript{2}Res Block structure contains multiple levels of multiple residual learning branches to exploit the \textit{cardinality} characteristic of networks. Figure \ref{fig:mul2res2} illustrates the Mul\textsuperscript{2}Res Block structure used in CVEGAN, which has four residual learning branches at two different levels. The number of branches and levels can be adapted for different applications based on the computational resources available.

At the first level, the input of the Mul\textsuperscript{2}Res block is  fed into four residual learning branches. Each branch has a convolutional layer with various kernel sizes (1$\times$1, 3$\times$3, 5$\times$5 and 7$\times$7). This diversifies the feature maps extracted by the convolutional layers using various receptive field sizes. The ECBAM is also employed at both levels before the final skip connection.

\begin{figure}[ht]
\centering
\includegraphics[width=0.85\linewidth]{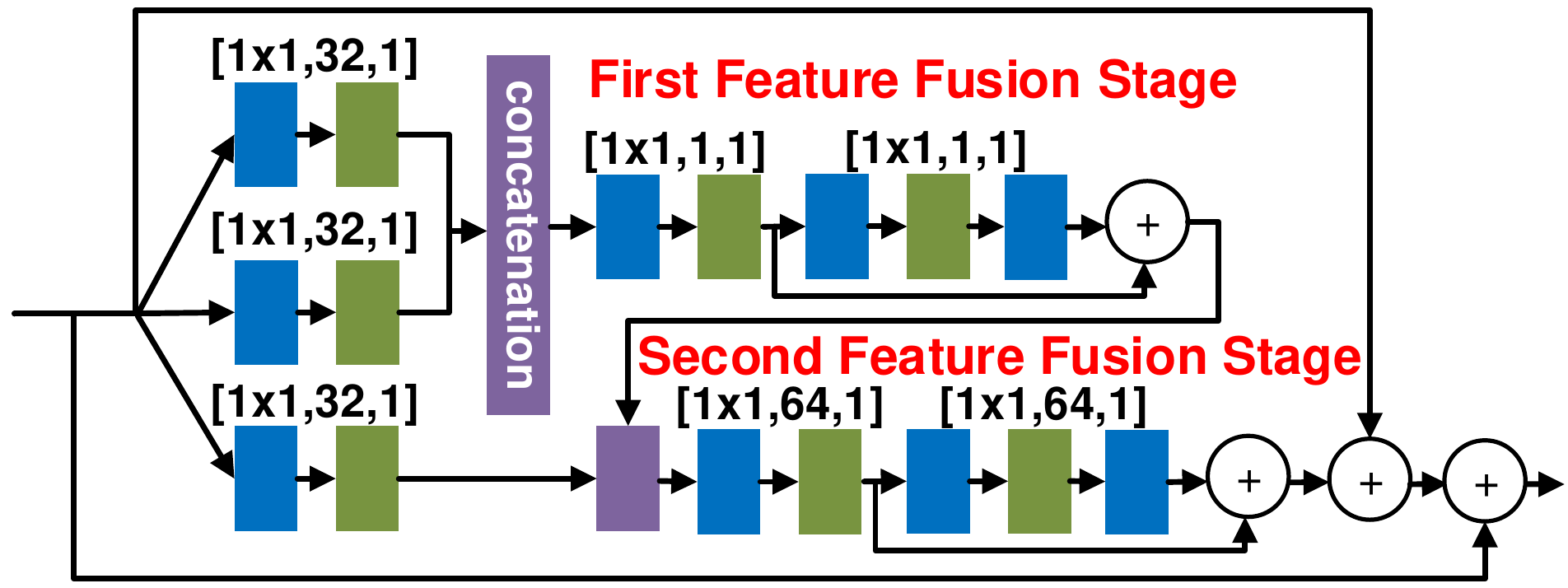}
\caption{Illustration of an ERNB.}
\label{fig:ernb}
\end{figure}

\begin{figure}[ht]
\centering
\includegraphics[width=0.85\linewidth]{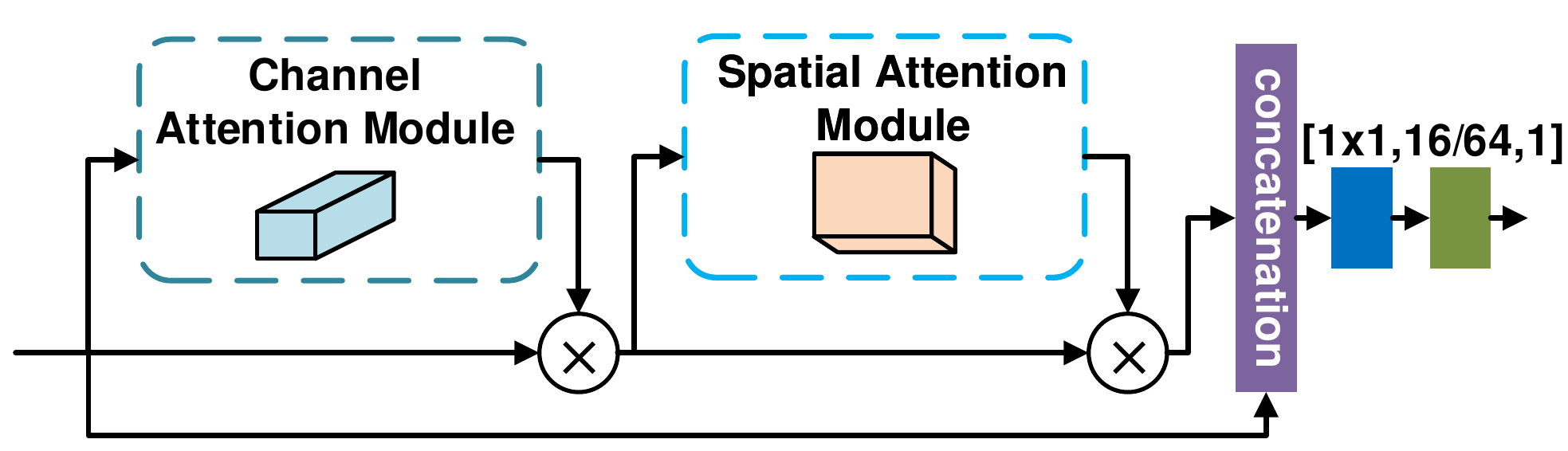}
\caption{Illustration of an ECBAM. \textcircled{$\times$}: matrix multiplication.}
\label{fig:ecbam}
\end{figure}

The Mul\textsuperscript{2}Res block at the second level also has four residual learning branches. The primary differences include (i) the Mul\textsuperscript{2}Res Blocks at the first level are replaced by residual blocks, each of which contains two convolutional layers (with various kernel sizes for different branches) and a Mish activation function; (ii) the output from four branches are concatenated before feeding into the ECBAM.

\textbf{ERNB}: Based on the non-local operations proposed in \cite{wang2018non,zhang2019nonlocal}, we have developed an enhanced residual non-local block (ERNB), as show in Figure \ref{fig:ernb}. This employs concatenation operations and residual blocks to achieve feature fusion, avoiding large matrix multiplication and facilitating training of the feature fusion process. A further modification is the introduction of a long skip connection employed to produce the ERNB output; this is designed to further stabilise the non-local learning.

\begin{figure*}[ht]
\centering
\includegraphics[width=1.0\linewidth]{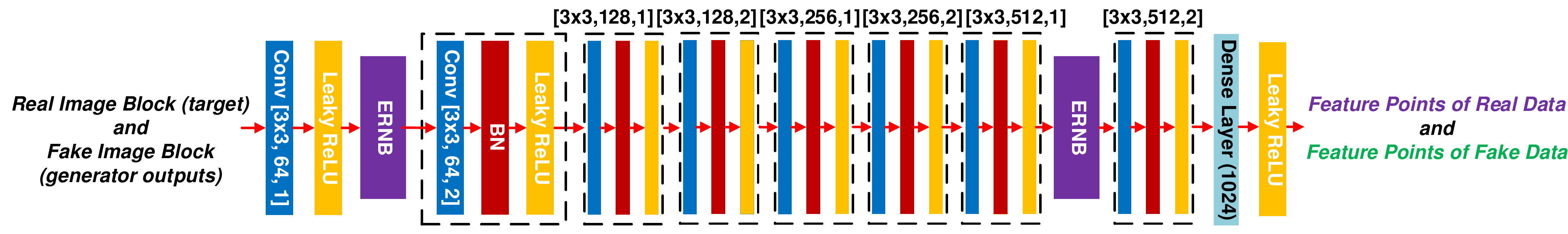}
\caption{Illustration of the CVEGAN's Discriminator.}
\label{fig:cvegan}
\end{figure*}

\textbf{ECBAM}: Inspired by attention mechanisms proposed recently \cite{zhang2018image,woo2018cbam,zhang2019nonlocal}, we have designed an enhanced convolutional block attention module (ECBAM) for CVEGAN (Figure \ref{fig:ecbam}). This follows the basic structure of the convolutional block attention module (CBAM) \cite{woo2018cbam}, which comprises a channel and a spatial attention module. Rather than directly producing output from the second matrix multiplication as in \cite{woo2018cbam}, we have added a concatenation operation with a convolutional layer to achieve non-linear feature fusion. This has been previously reported to improve information flow and overall performance of the network \cite{zhang2018residual}.

\subsection{Discriminator Architecture}

Figure \ref{fig:cvegan} illustrates the architecture of the CVEGAN discriminator, which is based on SRGAN \cite{ledig2017photo}. The primary differences include (i) we employ two ERNBs - one after the first convolutional layer, and the other before the final convolutional layer - these extract non-local features and improve the representational capability. (ii) we remove the final dense layer in SRGAN to output high (1024) dimensional feature points  rather than a single 1D scalar.

\section{Training Methodology}

We follow a similar training strategy to that described in \cite{ledig2017photo,wang2018esrgan}, which consists of two stages:  (i) CVENet is trained using a combined perceptual loss function (Section \ref{subsec:perceptualloss}) to obtain an initial model for the second training step; (ii) CVENet is then trained jointly with the discriminator with a new training method ReSphereGAN (Section \ref{subsec:respheregan}).

\subsection{Perceptually-inspired Loss Function}
\label{subsec:perceptualloss}

As discussed in Section \ref{subsec:gan}, existing loss functions combining pixel-wise losses, feature map differences and low-complexity quality metrics, do not always correlate well with perceived quality. To address this we have employed a loss function comprising a linear combination of the elementary transforms of six commonly used losses,  $\ell 1$ (denoted as $L_1$), $\ell 2$ ($L_2$), gradient loss \cite{cai2020rcagan,muhammad2020deep} ($L_3$), VGG19-54 loss \cite{wang2018esrgan} ($L_4$), Structural Similarity Index (SSIM) loss  \cite{wang2004image} ($L_5$) and Multi-scale SSIM (MS-SSIM) loss \cite{wang2003multiscale} ($L_6$): 
\small
\begin{equation}
L_\mathrm{test}=\sum_{i=1}^{6}{a_if(L_i)}.
\label{eq:loss}
\end{equation}
\normalsize
Here $f(\cdot)$ represents an elementary transformation \cite{rudin1964principles}, which can be either a constant, a power, a root, an exponential, a logarithmic, a trigonometric, an inverse trigonometric, a hyperbolic or an inverse hyperbolic function. $a_i$ represents the linear combination weights, where $a_6$ is always set to 1. The range of all these six single losses is between 0 and 1. We have excluded non-linear combinations and combinations of different transformed losses due to their high computational and training complexity.

We have used an eight-fold cross validation method \cite{howell2012statistical} to train the proposed combined loss function (equation (\ref{eq:loss})) based on eight publicly available subjective video quality databases. These include: the Netflix public database (70 test sequences) \cite{netflixpublic}, BVI-HD (192) \cite{zhang2018bvi}, CC-HD (108) \cite{zhang2020comparing}, CC-HDDO (90) \cite{katsenou2019subjective}, MCL-V (96) \cite{lin2015mcl}, SHVC (32) \cite{shvc}, IVP (100) \cite{ivp}, and VQEG-HD3 (72) \cite{vqeg}. All of these contain video sequences compressed using commonly used video codecs  (H.264, HEVC, AV1, VVC or MPEG-2). 

The eight databases were divided into two sub groups - seven training datasets and one for testing. An exhaustive search was performed among all tested transformation functions and their corresponding weighting parameters, the best of which was selected for this split to achieve the highest average correlation between the combined loss values and subjective scores for all seven training datasets. The Spearman rank order correlation coefficient (SROCC) is employed to quantify the correlation performance. The search range of each parameter is between 0 and 1, with an interval of 0.1.

\begin{table}[ht]
\centering
\scriptsize
\caption{Cross-validation results over eight training-testing trails.}
\begin{tabular}{l| M{0.5cm}|M{0.5cm}|M{0.5cm}|M{0.5cm}|M{0.5cm}|r}

\toprule
\multirow{1}{*}{Loss Function} & \multicolumn{1}{c|}{$L_{1}$} & \multicolumn{1}{c|}{$L_{2}$}& \multicolumn{1}{c|}{$L_{3}$}& \multicolumn{1}{c|}{$L_{4}$}& \multicolumn{1}{c|}{$L_{5}$}& \multicolumn{1}{c}{$L_{6}$}\\
\midrule
 Average SROCC&0.5984&0.6478&0.3991&0.7085&0.5720&0.7168 \\
 \midrule \midrule
 \multirow{1}{*}{Loss Function} & \multicolumn{1}{c|}{$L_{7}$} & \multicolumn{1}{c|}{$L_{8}$}& \multicolumn{1}{c|}{$L_{9}$}& \multicolumn{1}{c|}{$L_{10}$}& \multicolumn{1}{c|}{$L_{11}$}& \multicolumn{1}{c}{Ours}\\
\midrule
 Average SROCC&0.5430&0.5993&0.6008&0.6198&0.6591&\textbf{0.8067}
\\\bottomrule
\end{tabular}
\label{tab:srocc}
\end{table}

To avoid possible content bias due to a single training-testing split, we performed this cross-validation for all eight splits.  Table \ref{tab:srocc} presents the average SROCC performance on the test datasets among all eight splits for the trained loss functions, which are compared to the results from 11 commonly used loss functions in training image restoration and enhancement CNNs, $L_1$-$L_6$; linear combination of $\ell 1$ loss, gradient loss and VGG19-54 perceptual loss  \cite{muhammad2020srrescgan} ($L_7$); linear combination of $\ell 1$ loss and SSIM loss \cite{ma2019perceptually} ($L_8$); linear combination of $\ell 1$ loss and VGG19-54 perceptual loss \cite{wang2018esrgan,ahn2019photo,ren2020rcangan,ji2020patchesrgan,shang2020rfbesrgan} ($L_9$); linear combination of $\ell 1$ loss, gradient loss, SSIM and MS-SSIM losses and VGG19-54 perceptual loss \cite{cai2020rcagan} ($L_{10}$); linear combination of MSE and VGG19-54 perceptual loss \cite{ledig2017photo} ($L_{11}$). It can be observed that our trained loss functions have an average SROCC value of 0.8067, which is significantly higher than those for other tested loss functions ($L_1$-$L_{11}$).

The transformation function used for all optimal loss functions across all eight training-testing splits, is the natural logarithm $ln(\cdot)$. We use the median values of the corresponding combination parameters, and normalise these to ensure $\sum_{i=1}^6{a_i}=1$. The final combined loss function $ \mathcal L_{P}$ used to train CVEGAN is given below:
\begin{equation}
\begin{array}{cl}
        \mathcal L_{P}=& 0.3 \cdot ln (\ell 1)  + 0.2 \cdot  ln(\mathrm{SSIM}\_\mathrm{loss}) +\\ 
        &  0.1 \cdot  ln(\ell 2)+ 0.4 \cdot ln(\mathrm{MSSSIM}\_\mathrm{loss})
        \end{array}
        \label{eq:perceptual}
\end{equation}

It should be noted that $ \mathcal L_{P}$ remains within the range  0 to 1, and is differentiable, which is required to support back-propagation during training. Since the test sequences were generated using a wide range of video codecs, this loss function should generalise well across image and video compression applications.  $\mathcal L_{P}$ has been used here for training the CVENet (generator) during the first training stage.

\subsection{ReSphereGAN}
\label{subsec:respheregan}

The proposed Relativistic SphereGAN training methodology is a modified version of the SphereGAN \cite{park2019sphere}, which is based on an \textit{integral probability metric} (IPM). As illustrated in Figure \ref{fig:respheregan}, the original SphereGAN compares the geodesic distances between the north pole $\mathbf{N}$ and fake/real feature points in the $n$-dimensional Euclidean feature space through an inverse stereographic projection. Here $n$ is the feature point number of fake and real data produced by the CVEGAN discriminator, which has a default value of 1024. Inspired by RaGAN \cite{jolicoeur2018relativistic}, we also calculate the relativistic geodesic distance between the projected real and fake feature points in our loss functions at the second training stage.  This modification further optimises the generator by obtaining gradient information from both real and fake data during the adversarial training process. Specifically, the loss functions for generator ($\mathcal L_{Re\_gen}$) and discriminator ($\mathcal L_{Re\_disc}$) (in the second training stage) are given below:

\begin{figure}[ht]
\centering
\includegraphics[width=0.95\linewidth]{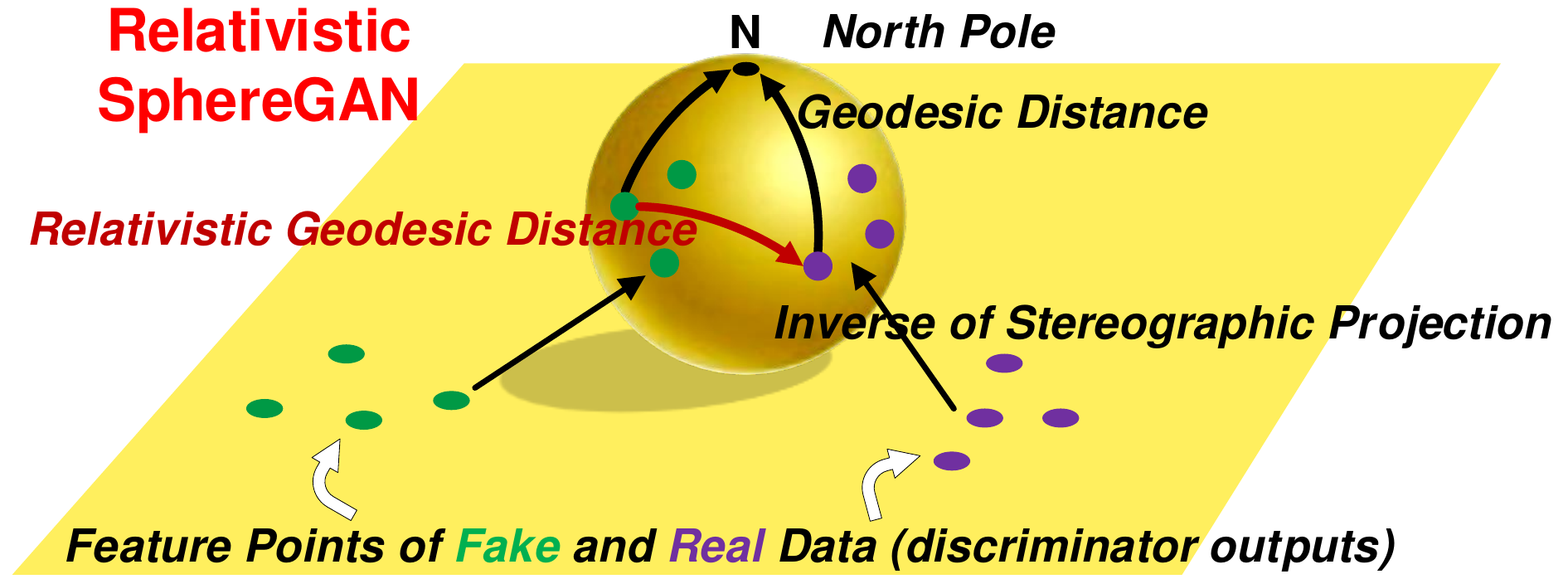}
\caption{Illustration of the proposed ReSphereGAN. Yellow plane and sphere represent the 1024-dimensional Euclidean feature space and hypersphere respectively. Green and purple points represent the feature points of fake and real data respectively.}
\label{fig:respheregan}
\end{figure}

\begin{equation}
\begin{array}{ll}
	\mathcal L_{Re\_gen}=&\mathcal L_{P} + 0.005 \cdot (-\sum _{m=1}^{M} E(d^m(\mathbf {N},T(\mathbf{x_f}))) \\ 
	&+ \sum _{m=1}^{M} E(d^m(T(\mathbf{x_r}),T(\mathbf{x_f}))))
	\end{array}
\label{eq:ResphereGAN1}
\end{equation}
\begin{equation}
\begin{array}{l}
	\mathcal L_{Re\_disc}=\sum _{m=1}^{M} E(d^m(\mathbf {N},T(\mathbf{x_f}))) -\\
	\sum _{m=1}^{M} E(d^m(\mathbf {N},T(\mathbf{x_r}))) - \sum _{m=1}^{M} E(d^m(T(\mathbf{x_r}),T(\mathbf{x_f})))
		\end{array}
\label{eq:ResphereGAN2}
\end{equation}
\normalsize
Here , $E(\cdot)$ represents the mean operation. $d^m(\mathbf{a},\mathbf{b})$ is the \textit{geometric-aware transformation function} \cite{park2019sphere}, which calculates the $m$-\textit{th} central geometric moment (geodesic distance) \cite{park2019sphere} in the hypersphere space between $\mathbf{a}$ and $\mathbf{b}$ (projected feature points in the hypersphere space). $M$ is set to 3 in this work. $T(\cdot)$ stands for the \textit{inverse of stereographic projection} \cite{park2019sphere} from the Euclidean feature space to the hypersphere space.

$\mathbf {N}$ is denoted as the north pole in the hypersphere space, and $\mathbf{x_r}$ and $\mathbf{x_f}$ are the real and fake feature points respectively in $n$-dimensional Euclidean feature space. The weight combining the perceptual loss ($\mathcal L_{P}$) and the generator adversarial loss was set to 0.005 based on several previous works \cite{wang2018esrgan,ma2019perceptually,shang2020rfbesrgan} on GAN-based image restoration and enhancement.

\subsubsection*{Gradient Analysis}

The gradients generated from the losses during the second training stage are crucial for stabilising the GAN models. It is important to avoid vanishing gradients and explosion problems as pointed out in \cite{goodfellow2014generative,goodfellow2016deep}. The gradients of $d^m(\mathbf{N},T(\mathbf{x_r}))$ and $d^m(\mathbf{N},T(\mathbf{x_f}))$ have already been evaluated in \cite{park2019sphere}, so we just  analyse the gradients of the new relativistic geodesic distance $d^m(T(\mathbf{x_r}),T(\mathbf{x_f}))$. 

\vspace{8pt}
\textbf{Lemma 1.} $E(\left\|\nabla_{(T(\mathbf{x_r}),T(\mathbf{x_f}))} d^m(T(\mathbf{x_r}),T(\mathbf{x_f}))\right\|_2)<\infty$ for all $m$ ($\nabla$ represents the derivation operation, and $\left\|\cdot \right\|_2$ is the Euclidean norm).
\vspace{8pt}

The proof of \textbf{Lemma 1} is provided in the \textbf{Supplementary Material}. This indicates that the proposed ReSphereGAN will ensure stable GAN learning with any moment $m$.

\section{Experiment Configuration}

\begin{figure}[ht]
\centering
\includegraphics[width=0.85\linewidth]{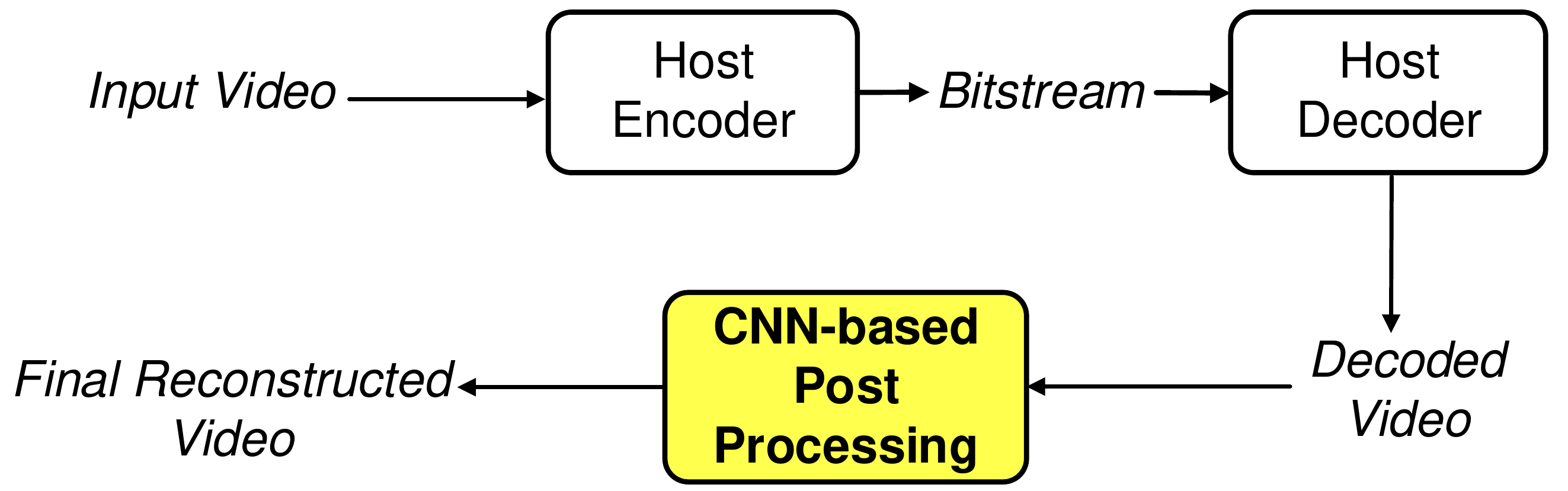}
\caption{Coding workflow with a CNN-based PP module.}
\label{fig:pp}
\end{figure}

As mentioned in Section \ref{subsec:dvc}, we have trained and evaluated the proposed network for two coding tools: post-processing (PP) and spatial resolution adaptation (SRA). Other important tools for compression enhancement, e.g. in-loop filtering and effective bit depth adaptation, have similar workflows to PP and SRA but offer lower overall coding gains \cite{zhang2019vistra2,JVET-O0063}. These are therefore not presented here due to the limited space available. 

Although the first version of VVC has recently been approved by standardisation bodies, we chose to use HEVC Test Model HM 16.20 as the host codec here as proof of concept. This is because HEVC has a low complexity compared to VVC and is hence more suitable for the large scale experimentation employed in this study. Based on our previous investigations, we anticipate that the gains for other codecs (AV1, VVC) will be commensurate with those presented for HEVC. However this remains for future work.  

\subsection{CNN-based PP and SRA Coding Tools}

\begin{figure}[ht]
\centering
\includegraphics[width=0.85\linewidth]{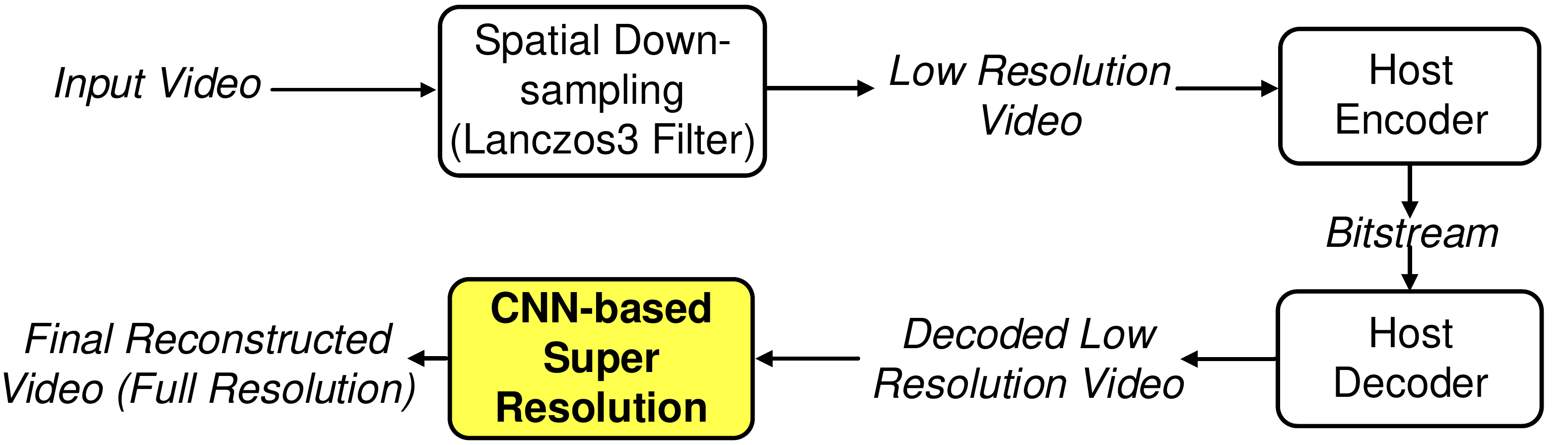}
\caption{Coding workflow with a CNN-based SRA module.}
\label{fig:sra}
\end{figure}

The coding workflows for CNN-based PP and SRA tools are shown in Figure \ref{fig:pp} and \ref{fig:sra}. The CNN module employed for PP provides an additional quality enhancement on the decoded content, while for SRA, the original video is spatially down-sampled (by a factor of 2) before encoding and a CNN-based super resolution operation is performed on the decoded low-resolution content for both up-sampling and quality enhancement. More detailed descriptions of PP and SRA can be found in \cite{ma2020bvi}.

\subsection{Training Data and Configuration}

A large training database, BVI-DVC \cite{ma2020bvi}, has been employed  to train the proposed network. This contains 800 video clips (10 bit, YCbCr 4:2:0) at various spatial resolutions from 270p to 2160p covering a wide range of content and video texture types. It has been previously reported to provide enhanced training performance over other databases for optimising CNN-based coding enhancement tools. It has recently been adopted as the default training database by JVET Ad-hoc Group 11 (Neural-network-based video coding) \cite{JVETT0042}.

We followed the same procedure as detailed in \cite{ma2020bvi} to generate training material for CNN-based PP and SRA coding tools using the HEVC HM 16.20 with the Random Access (RA) configuration (Main10 profile) and four base quantisation parameter (QP) values, 22, 27, 32 and 37. This produces two classes of training data for PP and SRA, each of which contains four QP sub-groups. For the SRA class, the nearest neighbour (NN) filter was utilised to up-sample the compressed low resolution video frames to the original resolution, which  ensures that the same CVENet architecture can be used for both PP and SRA.

For each QP sub-group in the PP and SRA classes, the compressed (or the NN up-sampled compressed) video frames and their corresponding original counterparts were randomly selected, cropped into 96$\times$96 blocks and converted YCbCr 4:4:4 format. Finally, approx. 195,000 pairs of image blocks for each QP sub-group were generated.

CVEGAN was implemented using the TensorFlow platform (1.8.0). The training process was conducted based on the following parameters: Adam optimisation \cite{kingma2014adam} with the following hyper-parameters:  $\beta_1$=0.9 and $\beta_2$=0.999; batch size of 16; 200 training epochs; initial learning rate (0.0001); weight decay of 0.1 for every 100 epochs.

\subsection{Evaluation Datasets and Configuration}

\begin{table*}[ht]
\scriptsize
\centering
\caption{Compression results in terms of BD-rate (based on both PSNR and VMAF) of the proposed CVEGAN, its ablation study variants and a selection of benchmark networks in the context of PP and SRA. All of them are benchmarked on the original HEVC - negative BD-rates indicate coding gains. The result sets \{i/j/k\} in this table stands for the BD-rate values for JVET-CTC, UVG and o-1-f respectively.}
\begin{tabular}{l ||M{1.7cm}|M{1.7cm}|M{1.65cm}||M{1.7cm}|M{1.7cm}|M{1.65cm}}
\toprule
\multirow{2}{*}{CNN/GAN Model} & \multicolumn{3}{c||}{CNN-based Post-Processing}&
\multicolumn{3}{c}{CNN-based Spatial Resolution Adaptation}\\
\cmidrule{2-7}
\centering
\scriptsize
&   BD-rate (\%)  &    BD-rate(\%)  & Relative  &BD-rate (\%)&    BD-rate (\%) & Relative \\
&  (PSNR)  &  (VMAF)  &Complexity&(PSNR)&    (VMAF) & Complexity\\
\midrule \midrule
SRCNN \cite{dong2015image}&-1.9/--/-- & -7.4/--/-- &1.0$\times$ &-3.1/--/-- & -21.1/--/-- & 1.0$\times$ \\
UDSR\cite{cai2019ntire}&-11.4/--/-- & -16.0/--/-- & 3.04$\times$&-15.7/--/-- & -33.3/--/-- & 5.62$\times$ \\
RNAN \cite{zhang2019nonlocal}&-12.5/-14.1/-11.7& -19.2/-23.9/-21.5& 5.78$\times$& -17.4/-9.3/--& -34.8/-31.6/--& 10.79$\times$ \\
MSRGAN \cite{ma2019perceptually}& -6.5/-8.7/-5.9& -21.1/-25.7/-23.5& 2.46$\times$& -9.1/-3.3/--& -35.6/-32.9/--& 4.54$\times$\\

RFB-ESRGAN \cite{shang2020rfbesrgan}& -9.1/-10.8/-7.9& -18.3/-23.2/-20.4&4.58$\times$& -12.9/-4.5/--& -34.3/-31.0/--&8.52$\times$ \\

\midrule
CVEGAN (Ours)&-10.2/-11.9/-9.0 & \textbf{-23.4/-27.8/-25.3} & 2.80$\times$ &-14.8/-6.4/--& \textbf{-38.4/-35.5/--} &5.23$\times$ \\
\midrule
CVEGAN (ResNeSt block \cite{zhang2020resnest}) &-9.4/--/-- & -22.3/--/-- & 2.81$\times$ &-14.3/--/--& -37.1/--/-- &5.25$\times$ \\
CVEGAN (Non-local block \cite{zhang2019nonlocal}) &-9.9/--/-- & -22.1/--/-- & 3.22$\times$ &-14.5/--/--& -37.2/--/-- & 6.20$\times$ \\
CVEGAN (CBAM block \cite{woo2018cbam}) &-10.0/--/-- & -22.3/--/-- & 2.78$\times$ &-14.2/--/--& -37.0/--/-- &5.13$\times$  \\
CVEGAN (SphereGAN \cite{park2019sphere}) &-9.7/--/-- & -22.1/--/-- & 2.80$\times$ &-14.3/--/--& -37.1/--/-- &5.23$\times$ \\
CVEGAN ($L_8$ loss)&-7.5/--/-- & -21.9/--/-- & 2.80$\times$ &-12.4/--/--& -37.0/--/-- &5.23$\times$ \\
\bottomrule
\end{tabular}
\label{tab:24cnn}
	\end{table*}

During evaluation, each decoded frame (after NN up-sampling for SRA) was converted to YCbCr 4:4:4 format and segmented into 96$\times$96 overlapping blocks with an overlap size of 4 pixels as CVENet (the generator) input. The network output blocks from CVENet were further aggregated following the same pattern and converted to YCbCr 4:2:0 format to form the final reconstructed frame.

Twenty-four popular and state-of-the-art CNN and GAN architectures, which have been widely used in image super-resolution, restoration and video compression, have been benchmarked in this paper. All of these have been re-implemented and trained using the same framework (TensorFlow 1.8.0) with identical training material following the same training methodology and loss functions as described in their original literature. During re-implementation, the input and output interfaces of these networks have been modified to satisfy the data format requirements.

All networks under test have been integrated into both PP and SRA coding tools and tested under JVET common test conditions (CTC) {\cite{jvetctc}} using the RA configuration (Main10) with four QP values of 22, 27, 32 and 37. The original HEVC HM 16.20 was used as the host codec and also as the benchmark anchor. The coding performance against HEVC HM is calculated using the Bj{\o}ntegaard Delta  {\cite{BD}} measurement (BD-rate) based on two quality metrics: Peak Signal-to-Noise-Ratio (PSNR, luma channel only) and Video Multimethod Assessment Fusion (VMAF, 0.6.1) {\cite{li2016toward}}. PSNR is the most widely used quality metric for image and video compression, although it does not always correlate well with subjective quality scores \mbox{\cite{sheikh2005information,zhang2018bvi}}. VMAF is a machine learning-based video quality assessment algorithm, increasingly used in industry, which combines various quality metrics and video features using a Support Vector Machine regressor. It has been reported to offer much better correlation performance with subjective opinions on compressed content compared to the PSNR \mbox{\cite{li2016toward,zhang2018bvi,bampis2018spatiotemporal}}.

The relative computational complexity of all test networks (the generator for the case of GANs) have also been calculated and benchmarked against the simplest SRCNN {\cite{dong2015image}}. The training and evaluation processes were both executed on a shared cluster, where each node contains two 14 core 2.4 GHz Intel E5-2680 V4 (Broadwell) CPUs, 128 GB of RAM, and NVIDIA P100 GPU devices.

To evaluate their performance, JVET-CTC SDR (19 test sequences) {\cite{jvetctc}} has been employed as the main test dataset due to its content and resolution diversity. The average BD-rates assessed by both PSNR and VMAF were calculated for all tested networks. To further evaluate network generalisation, another two commonly used test databases, UVG (6 test sequences) {\cite{mercat2020uvg}} and AOM main test dataset (21 test sequences) \textit{objective-1-fast} (o-1-f) \footnote{We have excluded a few source sequences in UVG and o-1-f datasets, which have been employed in BVI-DVC as training data.} {\cite{chen2020overview}}, have also been employed to test the proposed CVEGAN and the other three top performers. None of the sequences in these three test datasets were included in the training database, BVI-DVC. It should be noted that only UHD (2160p) content from these databases was used to evaluate the SRA coding tool since, as previously reported {\cite{zhang2019vistra2}}, lower resolutions  provide only limited and inconsistent coding gains.

We have also conducted a lab-based subjective test using a double stimulus methodology on a selection of network architectures (alongside the anchor HM and CVEGAN) for both PP and SRA. We collected subjective scores from 10 subjects \footnote{Due to the impact of the COVID-19 pandemic, a large lab-based subjective test employing non-expert participants could not be conducted.} using the reconstructed videos of 12 UHD source sequences (QP 37 only). Further details of the testing configuration can be found in the \textbf{Supplementary Material}.

\subsection{Ablation Study}

Five primary contributions have been tested and compared to the state of the art for PP and SRA. All ablation studies are based on the JVET-CTC SDR dataset. 

(1) \textbf{Mul\textsuperscript{2}Res Block} effectiveness has been evaluated by replacing it with other commonly used convolutional blocks for CNN-based image restoration, which include residual block (RB) \cite{ledig2017photo}, modified residual block (MRB) \cite{ma2019perceptually}, residual dense block (RDB) \cite{zhang2018residual}, residual-in-residual dense block (RRDB) \cite{wang2018esrgan}, residual channel attention block (RCAB) \cite{zhang2018image}, Xception block \cite{chollet2017xception} (\textit{cardinality} is 4), ResNeXt block \cite{xie2017aggregated} (\textit{cardinality} is 4) and ResNeSt block \cite{zhang2020resnest} (\textit{cardinality} is 4).

(2) \textbf{ERNB} is substituted by the original non-local block \cite{zhang2019nonlocal} to evaluate its effectiveness.

(3) \textbf{ECBAM} is replaced by the original CBAM \cite{woo2018cbam} for comparison.

(4) \textbf{ReSphereGAN} training has been compared with other commonly used GAN training approaches, including standard GAN \cite{ledig2017photo}, Relativistic average GAN (RaGAN) \cite{wang2018esrgan}, PatchGAN \cite{ji2020patchesrgan}, conditional GAN (cGAN) \cite{miyato2018cgans}, Wasserstein GAN-gradient penalty (WGAN-GP) \cite{gulrajani2017improved} and the original SphereGAN \cite{park2019sphere}.

(5) \textbf{Perceptual Loss Function} proposed in this paper was compared with other commonly used loss functions for GAN training ($L_7$-$L_{11}$),  as described in the Section \ref{subsec:perceptualloss}.

\section{Results and Discussion}

Table {\ref{tab:24cnn}} summarises the performance for CVEGAN and a selection of the top performing networks, alongside the ablation study results, with more comprehensive results provided in the \textbf{Supplementary Material}. It can be observed that, for both PP and SRA, the proposed CVEGAN outperforms all other tested architectures based on the perceptual quality metric VMAF. It also has a relatively low complexity compared to UDSR, RNAN and RFB-ESRGAN. From the ablation study, we observe that  the proposed Mul\textsuperscript{2}Res block, ERNB, ECBAM, ReSphereGAN and the new perceptual loss function have contributed at least 1.1\%, 1.2\%, 1.1\%, 1.3\% and 1.4\% coding gains (assessed by VMAF) when compared to the tested replacements, but with similar or lower complexity.

Table {\ref{tab:mos}} presents the average DMOS (difference of the mean opinion score) values of all evaluated sequences for CVEGAN, HEVC and the three top performing networks. The average DMOS for CVEGAN is lower than that for HEVC anchor and the other networks,  providing further evidence of its effectiveness. Figure {\ref{fig:perceptualcom}} also shows a subjective comparison between the output of HM 16.20, RNAN {\cite{zhang2019nonlocal}}, MSRGAN {\cite{ma2019perceptually}} and the proposed CVEGAN. More results can be found in the \textbf{Supplementary Material}.

\begin{table}[htbp]
\centering
\scriptsize
\caption{Subjective results based on 12 UHD source sequences.}
\centering
\begin{tabular}{l|M{1.97cm}| M{0.4cm}|M{0.4cm}|M{0.4cm}|M{0.4cm}}

\toprule
\multirow{1}{*}{PP} & Anchor (HM 16.20)& \multicolumn{1}{c|}{\cite{zhang2019nonlocal}} & \multicolumn{1}{c|}{\cite{ma2019perceptually}}& \multicolumn{1}{c|}{\cite{shang2020rfbesrgan}}& Ours\\
\midrule
 Average DMOS& 1.97&1.60& 1.57& 1.59&\textbf{1.53} \\ \midrule \midrule
 \multirow{1}{*}{SRA} &Anchor (HM 16.20)& \multicolumn{1}{c|}{ \cite{zhang2019nonlocal}} & \multicolumn{1}{c|}{ \cite{ma2019perceptually}}& \multicolumn{1}{c|}{ \cite{shang2020rfbesrgan}}& Ours\\
\midrule
 Average DMOS& 1.97&1.39& 1.41& 1.42& \textbf{1.34}
\\\bottomrule
\end{tabular}
\label{tab:mos}
\end{table}
\section{Conclusion}

In this paper, a novel GAN architecture, CVEGAN, has been proposed for compressed video quality enhancement. This network, when integrated into a conventional video coding system, has enabled significantly improved coding performance compared to many state-of-the-art architectures. This enhanced performance can be attributed to the use  of several new features including a novel Mul\textsuperscript{2}Res blocks, ERNB, ECBAM, the new ReSphereGAN training methodology and perceptual-inspired loss functions. Future work will address network complexity reduction and validation on other standardised codecs. We will also extend our approach to other application scenarios, including image restoration and super-resolution processing.  


\section{Acknowledgement}

We would like to acknowledge funding from UK EPSRC (EP/L016656/1 and EP/M000885/1) and the NVIDIA GPU Seeding Grants.

{\small
\bibliographystyle{IEEEtran}
\small\bibliography{IEEEabrv,egbib}
}

\section*{Supplementary Material}

The Supplementary Material is organised as follows. Section \ref{sec:proof} presents the mathematical proof of the \textbf{Lemma 1}. Section \ref{sec:results} provides the comprehensive comparison results between the proposed CVEGAN and 24 state-of-the-art networks, and full Ablation Study results. The subjective experiment configurations are described in the Section \ref{sec:subjective}. Finally, Section \ref{sec:comparisons} provides additional perceptual comparisons.

\subsection{Mathematical Proof of Lemma 1}
\label{sec:proof}

\textbf{Lemma 1.} $E(\left\|\nabla_{(\mathbf{x_r},\mathbf{x_f})} d^m(T(\mathbf{x_r}),T(\mathbf{x_f}))\right\|_2)<\infty$ for all $m$. Here $\nabla$ represents the derivation operation, and $\left\|\cdot \right\|_2$ is the Euclidean norm.

\begin{proof}
Based on the definition of geodesic distance in \cite{park2019sphere}, the relativistic geodesic distance $d^m(T(\mathbf{x_r}),T(\mathbf{x_f}))$ can be written as:

\small
\begin{equation}
\begin{array}{ll}
\multicolumn{2}{l}{d^m(T(\mathbf{x_r}),T(\mathbf{x_f}))}\\
\displaystyle =&\displaystyle\mathrm{arccos}^m\left(\frac{\left\|\mathbf{x_r} \right\|_2^2 \left\|\mathbf{x_f} \right\|_2^2-\left\|\mathbf{x_r} \right\|_2^2-\left\|\mathbf{x_f} \right\|_2^2+4\mathbf{x_r}\cdot\mathbf{x_f}+1}{(\left\|\mathbf{x_r} \right\|_2^2+1)(\left\|\mathbf{x_f} \right\|_2^2+1)}\right)\\
 
 \displaystyle \equiv&\displaystyle\mathrm{arccos}^m(A)
        \label{eq:distance}
        \end{array}
\end{equation}
\normalsize

Here $A\equiv\displaystyle \frac{\left\|\mathbf{x_r} \right\|_2^2 \left\|\mathbf{x_f} \right\|_2^2-\left\|\mathbf{x_r} \right\|_2^2-\left\|\mathbf{x_f} \right\|_2^2+4\mathbf{x_r}\cdot\mathbf{x_f}+1}{(\left\|\mathbf{x_r} \right\|_2^2+1)(\left\|\mathbf{x_f} \right\|_2^2+1)}$, $A \in [-1, 1]$. $\displaystyle\mathbf{x_r}$ and $\displaystyle\mathbf{x_f}$ are the real and fake feature points respectively in $n$-dimensional Euclidean feature space.

According to the chain rule,  
\begin{equation}
\displaystyle \frac{\partial d^m(T(\mathbf{x_r}),T(\mathbf{x_f}))}{\partial \mathbf{x_r}}
=\displaystyle \mathrm{arccos}^{m-1}(A) \cdot
\frac{-m}{\sqrt{1-A^2}}\cdot \frac{\partial A}{\partial \mathbf{x_r}}
\label{eq:firststep}
\end{equation}

Based on the equation (\ref{eq:firststep}), the gradient of $d^m(T(\mathbf{x_r}),T(\mathbf{x_f}))$ can be further obtained following the chain rule and the product rule of derivative:

\begin{equation}
\begin{array}{ll}
\multicolumn{2}{l}{ \displaystyle\nabla_{(\mathbf{x_r},\mathbf{x_f})}d^m(T(\mathbf{x_r}),T(\mathbf{x_f}))}\\ 
=&\displaystyle \frac{\partial ^2d^m(T(\mathbf{x_r}),T(\mathbf{x_f}))}{\partial \mathbf{x_r}\partial \mathbf{x_f}}\\

=&\displaystyle m(m-1)\cdot\mathrm{arccos}^{m-2}(A) \cdot
\frac{1}{1-A^2}\cdot\frac{\partial A}{\partial \mathbf{x_f}}\cdot\frac{\partial A}{\partial \mathbf{x_r}} -\\

&  \displaystyle m\cdot\mathrm{arccos}^{m-1}(A) \cdot
\frac{A}{(1-A^2)^{\frac{3}{2}}}\cdot\frac{\partial A}{\partial \mathbf{x_f}}\cdot\frac{\partial A}{\partial \mathbf{x_r}} -\\
& \displaystyle m\cdot\mathrm{arccos}^{m-1}(A) \cdot \displaystyle\frac{1}{\sqrt{1-A^2}}\cdot\frac{\partial ^2A}{\partial \mathbf{x_r}\partial \mathbf{x_f}}
        \end{array}
        \label{eq:proof}
\end{equation}

Based on Lemma 1 and 2 in \cite{park2019sphere}, we have

\begin{equation}
\left\{
\begin{array}{l}
\displaystyle \mathrm{arccos}(A)<\infty\\
\displaystyle \frac{\partial A}{\partial \mathbf{x_f}}\frac{\partial A}{\partial \mathbf{x_r}}<\infty\\
\displaystyle  \frac{\partial ^2A}{\partial \mathbf{x_r}\partial \mathbf{x_f}}<\infty
\end{array}
\right.
\label{eq:4}
\end{equation}

According to Proposition 1 and 2 in \cite{park2019sphere} and Theorem 6.9 in \cite{villani2008optimal}, the geometric distance between the real and fake data feature points, $\mathbf{x_r}$ and $\mathbf{x_f}$, weakly converges to 0, for all moment values $m$:
\begin{equation}
    \displaystyle d^m(T(\mathbf{x_r}),T(\mathbf{x_f}))=\mathrm{arccos}^m(A) \rightharpoonup 0
\end{equation}
Here, $\rightharpoonup$ represents \textit{weak convergence}. This indicates that $\mathrm{arccos}^m(A) \neq 0$, and thus $A \neq \pm 1$ \cite{rudin1964principles}. Therefore we can have:

\begin{equation}
\displaystyle 1-A^2\neq 0
\label{eq:6}
\end{equation}

According to equation (\ref{eq:4}) and (\ref{eq:6}), the gradient (and its mean) of the relativistic geodesic distance is bounded for all moment values $m$:
\begin{equation}
    \nabla_{(\mathbf{x_r},\mathbf{x_f})}d^m(T(\mathbf{x_r}),T(\mathbf{x_f}))<\infty,
\end{equation}

\begin{equation}
    E(\left\|\nabla_{(\mathbf{x_r},\mathbf{x_f})} d^m(T(\mathbf{x_r}),T(\mathbf{x_f}))\right\|_2)<\infty
\end{equation}

\end{proof}

\textbf{Lemma 1} indicates that the ReSphereGAN can ensure stable GAN learning with any moment $m$. In practice, we have noted that although ReSphereGAN may generate relatively large gradients as the original SphereGAN,  they can still be calculated during training process when the Adam optimiser is used \cite{kingma2014adam}. This has also been reported in \cite{park2019sphere}.

\subsection{Experiments}
\label{sec:results}

In this section, comprehensive experimental results are presented including comparisons between the proposed CVEGAN and 24 state-of-the-art network architectures and Ablation Study analysis.

\subsubsection{Comprehensive Comparisons Results}

\begin{table*}[ht]
\small
\centering
\caption{Comprehensive compression results (in terms of BD-rate based on both PSNR and VMAF) of the proposed CVEGAN, its ablation study variants and 24 benchmark networks when they are integrated into PP and SRA coding tools for HEVC compression. All of them are benchmarked on the original HEVC - negative BD-rates indicate coding gains. The result sets \{i/j/k\} in this table stands for the BD-rate values for JVET-CTC, UVG and o-1-f respectively. The relative complexity of each test network is also provided for comparison.}
Compression performance comparisons with state-of-the-art network architectures\\
\scriptsize
\begin{tabular}{r ||M{1.7cm}|M{1.7cm}|M{1.65cm}||M{1.5cm}|M{1.7cm}|M{1.65cm}}
\toprule
\multirow{3}{*}{Network} & \multicolumn{3}{c||}{CNN-based Post-Processing}&
\multicolumn{3}{c}{CNN-based Spatial Resolution Adaptation}\\
\cmidrule{2-7}
\centering
&   BD-rate (\%)  &    BD-rate (\%)  & Relative  &BD-rate (\%)&    BD-rate (\%) & Relative \\
&  (PSNR)  &  (VMAF)  &Complexity&(PSNR)&    (VMAF) & Complexity\\
\midrule \midrule
SRCNN \cite{dong2015image}&-1.9/--/-- & -7.4/--/-- &1.0$\times$ &-3.1/--/-- & -21.1/--/-- & 1.0$\times$ \\
FSRCNN \cite{dong2016accelerating}&-1.6/--/-- & -7.3/--/-- &1.37$\times$&-4.5/--/-- & -20.9/--/-- & 1.28$\times$ \\
VDSR \cite{kim2016accurate}&-1.9/--/-- &-7.6/--/-- & 2.05$\times$&-6.6/--/-- & -18.3/--/-- & 3.79$\times$ \\
DRRN \cite{tai2017image}&-10.8/--/-- & -14.9/--/-- &2.70$\times$&-15.0/--/--& -33.2/--/-- & 5.01$\times$ \\
EDSR \cite{lim2017enhanced}&-10.0/--/-- & -14.6/--/-- &4.50$\times$&-13.4/--/-- & -30.1/--/-- & 8.33$\times$ \\
SRResNet \cite{ledig2017photo}&-9.8/--/-- & -12.7/--/-- &2.45$\times$&-13.2/--/-- & -30.0/--/-- & 4.46$\times$ \\
MSRResNet \cite{ma2019perceptually}&-10.4/--/-- & -14.2/--/-- & 2.46$\times$&-14.6/--/-- & -32.7/--/-- & 4.52$\times$ \\
CARN \cite{ahn2018fast}&-11.2/--/-- & -15.4/--/-- & 2.23$\times$&-15.5/--/-- & -33.5/--/-- & 4.15$\times$ \\
UDSR\cite{cai2019ntire}&-11.4/--/-- & -16.0/--/-- & 3.04$\times$&-15.7/--/-- & -33.3/--/-- & 5.62$\times$ \\
HR-EnhanceNet \cite{ignatov2019ntire}&-11.3/--/-- & -16.4/--/-- & 2.80$\times$ &-15.8/--/-- & -33.1/--/-- & 5.56$\times$\\
ESRResNet \cite{wang2018esrgan}&-11.8/--/-- & -17.7/--/-- &3.82$\times$&-16.1/--/-- & -33.6/--/-- & 7.10$\times$ \\
RCAN \cite{zhang2018image}&-12.1/--/-- & -18.5/--/-- & 4.82$\times$&-17.1/--/-- & -35.1/--/-- & 8.98$\times$ \\
RDN \cite{zhang2018residual}&-12.2/--/--& -17.0/--/-- & 3.46$\times$&-16.6/--/-- & -34.5/--/--& 6.41$\times$ \\
RNAN \cite{zhang2019nonlocal}&-12.5/-14.1/-11.7& -19.2/-23.9/-21.5& 5.78$\times$& -17.4/-9.3/--& -34.8/-31.6/--& 10.79$\times$ \\
ADGAN \cite{lin2019adgan}& -1.3/--/--& -7.7/--/--& 2.46$\times$& -5.1/--/--& -18.5/--/--&4.56$\times$ \\
SRResCGAN \cite{muhammad2020srrescgan}& -7.1/--/--& -10.4/--/--& 1.71$\times$& -10.3/--/--& -27.2/--/-- &3.16$\times$ \\
SRGAN \cite{ledig2017photo}& -7.4/--/--& -12.9/--/--& 2.46$\times$& -10.9/--/-- & -30.2/--/--&4.52$\times$ \\
PCARNGAN \cite{ahn2019photo}&-8.3/--/-- & -16.0/--/-- & 2.25$\times$& -12.3/--/--& -33.7/--/--&4.18$\times$ \\
RCAGAN \cite{cai2020rcagan}& -9.1/--/--& -16.8/--/--& 3.31$\times$& -13.7/--/--& -33.9/--/--&6.13$\times$ \\
MSRGAN \cite{ma2019perceptually}& -6.5/-8.7/-5.9& -21.1/-25.7/-23.5& 2.46$\times$& -9.1/-3.3/--& -35.6/-32.9/--& 4.54$\times$\\
ESRGAN \cite{wang2018esrgan}& -8.7/--/--& -17.9/--/--& 3.82$\times$& -12.5/--/--& -33.8/--/--&7.15$\times$ \\
RCAN-GAN \cite{ren2020rcangan}& -9.3/--/--& -18.0/--/--& 4.84$\times$& -13.9/--/--& -34.1/--/--&9.03$\times$ \\
PatchESRGAN \cite{ji2020patchesrgan}& -9.0/--/--& -18.1/--/--&3.82$\times$ & -12.8/--/--& -34.2/--/-- & 7.20$\times$ \\
RFB-ESRGAN \cite{shang2020rfbesrgan}& -9.1/-10.8/-7.9& -18.3/-23.2/-20.4&4.58$\times$& -12.9/-4.5/--& -34.3/-31.0/--&8.52$\times$ \\

\midrule
CVENet (Ours)&-9.5/-11.3/-8.5 & -21.3/-26.0/-23.6 & 2.80$\times$ &-14.2/-5.9/-- & -36.4/-33.3/-- &5.23$\times$ \\
CVEGAN (Ours)&-10.2/-11.9/-9.0 & \textbf{-23.4/-27.8/-25.3} & 2.80$\times$ &-14.8/-6.4/--& \textbf{-38.4/-35.5/--} &5.23$\times$ \\
\bottomrule
\end{tabular}
\vspace{10pt}

\small
Ablation Study Results
\scriptsize
\begin{tabular}{c || r ||M{1.7cm}|M{1.5cm}|M{1.65cm}||M{1.5cm}|M{1.7cm}|M{1.65cm}}
\toprule
\multicolumn{2}{c||}{\multirow{3}{*}{CVEGAN Variants} }& \multicolumn{3}{c||}{CNN-based Post-Processing}&
\multicolumn{3}{c}{CNN-based Spatial Resolution Adaptation}\\
\cmidrule{3-8}
\multicolumn{2}{c||}{}&   BD-rate (\%)  &    BD-rate (\%)  & Relative  &BD-rate (\%)&    BD-rate (\%) & Relative \\
\multicolumn{2}{c||}{}&  (PSNR)  &  (VMAF)  &Complexity&(PSNR)&    (VMAF) & Complexity\\
\midrule \midrule
 & w/ ResNeSt block \cite{zhang2020resnest} &-9.4/--/-- & -22.3/--/-- & 2.81$\times$ &-14.3/--/--& -37.1/--/-- &5.25$\times$ \\
& w/ RCAB block \cite{zhang2018image}&-9.0/--/-- & -22.0/--/-- & 3.07$\times$ &-14.0/--/--& -36.6/--/-- &5.72$\times$ \\
& w/ RRDB block \cite{wang2018esrgan}&-8.8/--/-- & -21.6/--/-- & 3.16$\times$ &-13.8/--/--& -36.3/--/-- &5.92$\times$ \\
w/o & w/ RDB block \cite{zhang2018residual}&-7.1/--/-- & -20.9/--/-- & 2.95$\times$ &-12.3/--/--& -35.9/--/-- &5.67$\times$ \\
Mul\textsuperscript{2}Res block& w/ ResNeXt block \cite{xie2017aggregated}&-6.7/--/-- & -20.6/--/-- & 2.65$\times$ &-12.0/--/--& -35.7/--/-- &4.85$\times$ \\
& w/ Xception block \cite{chollet2017xception}&-6.6/--/-- & -20.1/--/-- & 2.65$\times$ &-11.8/--/--& -35.2/--/-- &4.92$\times$ \\
& w/ MRB block \cite{ma2019perceptually}&-6.3/--/-- & -18.3/--/-- & 2.24$\times$ &-9.4/--/--& -34.6/--/-- &4.03$\times$ \\
& w/ RB block \cite{ledig2017photo}&-6.1/--/-- & -16.6/--/-- & 2.24$\times$ &-8.6/--/--& -33.1/--/-- &4.07$\times$ \\
\midrule
w/o ERNB& w/ Non-local block \cite{zhang2019nonlocal} &-9.9/--/-- & -22.1/--/-- & 3.22$\times$ &-14.5/--/--& -37.2/--/-- & 6.20$\times$ \\
\midrule
w/o ECBAM& w/ CBAM block \cite{woo2018cbam} &-10.0/--/-- & -22.3/--/-- & 2.78$\times$ &-14.2/--/--& -37.0/--/-- &5.13$\times$   \\
\midrule
& w/ SphereGAN \cite{park2019sphere} &-9.7/--/-- & -22.1/--/-- & 2.80$\times$ &-14.3/--/--& -37.1/--/-- &5.23$\times$ \\
& w/ RaGAN \cite{wang2018esrgan} &-9.6/--/-- & -21.8/--/-- & 2.80$\times$ &-14.2/--/--& -36.8/--/-- &5.26$\times$ \\
w/o & w/ cGAN \cite{miyato2018cgans} &-9.6/--/-- & -21.6/--/-- & 2.80$\times$ &-14.1/--/--& -36.7/--/-- &5.26$\times$ \\
ReSphereGAN& w/ PatchGAN \cite{ji2020patchesrgan} &-9.4/--/-- & -21.6/--/-- & 2.79$\times$ &-13.8/--/--& -36.8/--/-- &5.31$\times$ \\
& w/ WGAN-GP \cite{gulrajani2017improved} &-9.1/--/-- & -21.7/--/-- & 2.80$\times$ &-13.2/--/--& -36.6/--/-- &5.28$\times$ \\
& w/ Standard GAN \cite{ledig2017photo} &-8.2/--/-- & -21.5/--/-- & 2.80$\times$ &-12.6/--/--& -36.5/--/-- &5.26$\times$ \\
\midrule
& w/ $L_7$ loss \cite{muhammad2020srrescgan}&-4.5/--/-- & -19.2/--/-- & 2.80$\times$ &-9.6/--/--& -34.1/--/-- &5.21$\times$ \\
& w/ $L_8$ loss \cite{ma2019perceptually}&-7.5/--/-- & -21.9/--/-- & 2.80$\times$ &-12.4/--/--& -37.0/--/-- &5.23$\times$ \\
w/o $ \mathcal L_{P}$& w/ $L_{9}$ loss \cite{wang2018esrgan}&-5.7/--/-- & -20.2/--/-- & 2.80$\times$ &-10.7/--/--& -35.4/--/-- &5.21$\times$ \\
& w/ $L_{10}$ loss \cite{cai2020rcagan}&-6.4/--/-- & -21.4/--/-- & 2.80$\times$ &-11.5/--/--& -36.3/--/-- &5.26$\times$ \\
&w/ $L_{11}$ loss \cite{ledig2017photo}&-5.8/--/-- & -20.4/--/-- & 2.79$\times$ &-10.7/--/--& -35.5/--/-- &5.20$\times$ 
\\\bottomrule
\end{tabular}
\label{tab:24cnn1}
	\end{table*}

Twenty-four state-of-the-art network architectures, which have been widely used in image super-resolution, restoration and video compression, have been benchmarked in this paper. These include 14 typical CNN structures and 10 GAN architectures. All of these have been re-implemented and trained using the same framework (TensorFlow 1.8.0) with identical training material (BVI-DVC \cite{ma2020bvi}) following the same training methodology and loss functions as described in their original literature. Their names, original literature and primary features are summarised below:

\begin{itemize}
    \item \textbf{SRCNN} \cite{dong2015image} is the first CNN model designed for single image super resolution (SISR). It employs a simple network structure with only 3 convolutional layers.
    \item \textbf{FSRCNN} \cite{dong2016accelerating} was also developed for SISR, containing 8 convolutional layers with various kernel sizes. 
    \item \textbf{VDSR} \cite{kim2016accurate} contains 20 convolutional layers employing global residual learning.
    \item \textbf{SRResNet} \cite{ledig2017photo} was the first network structure with residual blocks designed for SISR.
    \item \textbf{DRRN} \cite{tai2017image} employs a recursive structure and also contains residual blocks. 
    \item \textbf{EDSR} \cite{lim2017enhanced} significantly increases the number of feature maps (256) for the convolutional layers in each residual block.
    \item \textbf{RDN} \cite{zhang2018residual} was the first network architecture to combine residual block and dense connections for SISR. 
    \item \textbf{ESRResNet} \cite{wang2018esrgan} enhances SRResNet by combining residual blocks with dense connections, and employs residual learning at multiple levels. 
    \item \textbf{RCAN} \cite{zhang2018image} incorporates a channel attention (CA) scheme in the CNN. 
    \item \textbf{MSRResNet} \cite{ma2019perceptually} modified SRResNet by removing the BN layers for all the residual blocks.
    \item \textbf{CARN} \cite{ahn2018fast} was the first network architecture to combine the cascading connections and residual blocks for SISR task.
    \item \textbf{UDSR} \cite{cai2019ntire} integrated the U-shape structure with deep residual learning to achieve improved performance for SISR.
    \item \textbf{HR-EnhanceNet} \cite{ignatov2019ntire} utilises the U-shape structure along with two modified HRNets \cite{sun2019deep}.
    \item \textbf{RNAN} \cite{zhang2019nonlocal} was the first network architecture to combine non-local operation \cite{wang2018non} with residual learning structures.
    \item \textbf{ADGAN} \cite{lin2019adgan} combines the U-shape structure with simple convolutional layers and trained the network using the standard GAN \cite{ledig2017photo}.
    \item \textbf{SRResCGAN} \cite{muhammad2020srrescgan} employs multiple residual blocks realise SISR based on the RaGAN training methodology \cite{wang2018esrgan}.
    \item \textbf{SRGAN} \cite{ledig2017photo} was the first network to combine the standard GAN training methodology with perceptual loss functions (VGG19 \cite{simonyan2014very}) for photo-realistic SISR.
    \item \textbf{PCARNGAN} \cite{ahn2019photo} employs cascading connections and residual learning blocks, and trained the network with the standard GAN methodology \cite{ledig2017photo}.
    \item \textbf{RCAGAN} \cite{cai2020rcagan} modified SRResNet by replacing the original residual blocks with residual channel attention blocks, and trained the network using the conditional GAN (cGAN) methodology \cite{miyato2018cgans}.
    \item \textbf{MSRGAN} \cite{ma2019perceptually} trained the modified SRGAN using the RaGAN methodology \cite{wang2018esrgan}.
    \item \textbf{ESRGAN} \cite{wang2018esrgan} employs the Relativistic GAN algorithm \cite{jolicoeur2018relativistic} to train the ESRResNet for SISR.
    
    \item \textbf{RCAN-GAN} \cite{ren2020rcangan} trained the RCAN \cite{zhang2018image} using both standard GAN \cite{ledig2017photo} and RaGAN training methodology \cite{wang2018esrgan}.
    
    \item \textbf{PatchESRGAN} \cite{ji2020patchesrgan} trained the ESRGAN \cite{wang2018esrgan} using the PatchGAN algorithm \cite{isola2017image}.
    
    \item \textbf{RFB-ESRGAN} \cite{shang2020rfbesrgan} modifies the ESRGAN \cite{wang2018esrgan} by replacing multiple residual-in-residual dense blocks with receptive field dense blocks (RFBs) and trained the network using the RaGAN methodology \cite{wang2018esrgan}.
\end{itemize}

Table {\ref{tab:24cnn1}} summarises the compression performance generated by CVEGAN and the 24 CNN/GAN networks when they are integrated into post-processing (PP) and spatial resolution adaptation (SRA) coding tools in the context of HEVC. We can observe that for both PP and SRA coding tools, although based on PSNR, the pixel-wise distortion based quality metric, the proposed CVEGAN is not the best performer among all the tested networks, it outperforms all 24 architectures based on the perceptual quality metric VMAF. Considering that VMAF offers much higher correlation with subjective scores compared to PSNR \cite{li2016toward,zhang2018bvi,bampis2018spatiotemporal}, the effectiveness of the proposed algorithm in term of video quality enhancement is evident. The additional coding gains in terms of BD-rate (based on VMAF) compared to other networks are greater than 1.8\% and 2.6\% for PP and SRA respectively. 

We also evaluated the models obtained after the first training step (trained with our perceptual loss function $ \mathcal L_{P}$), denoted as CVENet in Table \ref{tab:24cnn1}. Its overall performance is slightly lower than that of the final CVEGAN, with up to 2.1\% and 2.2\%  BD-rate differences (based on VMAF) for PP and SRA respectively. This demonstrates the improvement due to the second training stage using the proposed ReSphereGAN.

Table \ref{tab:24cnn1} also shows the relative complexity of all test networks, which are benchmarked on that of SRCNN. It is noted that the relative complexity of CVEGAN is only 2.8 times of that for SRCNN, which is relatively low compared to many network architectures including EDSR, UDSR, ESRResNet, RCAN, RDN, RNAN RCAGAN, ESRGAN, RCAN-GAN, PatchESRGAN and RFB-ESRGAN.

\subsubsection{Ablation Study}

As mentioned in our paper, five primary contributions of CVEGAN have been compared with multiple alternative structures. The full results are presented in Table \ref{tab:24cnn}. 
It can be observed that compared to each of the five primary features of CVEGAN, all its counterparts offer lower coding gains (based on VMAF) for both PP and SRA applications. This shows the effectiveness of these new structures.

\subsection{Subjective Test Configuration}
\label{sec:subjective}

We have conducted a lab-based subjective test on the results generated by CVEGAN and a selection of other tested network architectures, which have achieved relatively higher coding gains according to VMAF. Details on test content, experimental configurations and data processing are described in this section.

\subsubsection{Reference and Test Content}

Twelve UHD (2160p) source sequences from the JVET-CTC SDR (standard dynamic range) \cite{jvetctc} and UVG \cite{mercat2020uvg} datasets  are selected as source content in this subjective evaluation. They have been encoded by the original HEVC HM 16.20 and its two enhanced versions (QP 37 only) with CNN-based PP and SRA coding tools. Each tool further generated results using four different networks to perform CNN operations, including RNAN \cite{zhang2019nonlocal}, RFB-ESRGAN \cite{shang2020rfbesrgan}, MSRGAN \cite{ma2019perceptually}, and the proposed CVEGAN. The former three architectures were selected due to their relatively higher coding gains (see Table \ref{tab:24cnn}) compared to other benchmark networks assessed by VMAF. This results in 9 different test versions for each source sequence.




\subsubsection{Experimental Configurations}

The subjective tests were conduced in a laboratory with a darkened, living room style environment. The background luminance level was set to 15\% of the peak luminance of the monitor used \cite{subjectivetest}. All the video sequences were shown at their native framerates, on a SONY PVM-X550 4K OLED professional video monitor with a maximum viewing angle of 89$^{\circ}$ and an effective picture size (H$\times$V) of 1209.6$\times$680.4 mm. We connected this reference monitor (the spatial resolution was configured to 3840$\times$2160) to a Windows PC running the MATLAB R2019b and Psychotoolbox 3.0. The viewing distance was set to be 1.6 times of the monitor height (718.4 mm) based on the ITU-R BT.500 \cite{subjectivetest}.

In this experiments, the double stimulus continuous quality scale (DSCQS) methodology \cite{subjectivetest} was used. In each trial, participants were shown Sequence A and Sequence B twice. One of these is a source sequence and the other is one of its nine distorted versions. Their orders are randomly determined and unknown to the subjects.  After viewing these two sequences, the participants were asked to rate the perceived quality of both videos, based on a continuous quality scale from 1 to 5 (1-Bad, 2-Poor, 3-Fair, 4- Good and 5-Excellent).

Ten subjects (6 male and 4 female)\footnote{Due to the impact of the COVID-19 pandemic, a limited number of participants were employed in this experiment.} participated in this experiment and their average age was 31.6 years. They all had normal or corrected to-normal colour vision verified by using Snellen and Ishihara charts \cite{subjectivetest}. 

\subsubsection{Data Processing}

Difference scores were calculated for each trial and each participant by subtracting the quality score of the distorted sequence from its corresponding reference source. Possible outliers were removed following the procedures described in \cite{subjectivetest}. Difference mean opinion scores (DMOS) were obtained for every trial by taking the mean of the difference scores. We then calculated the average DMOS among all source sequences for each test version (HM 16.20, PP-RNAN, PP-RFB-ESRGAN, PP-MSRGAN, PP-CVEGAN, SRA-RNAN, SRA-RFB-ESRGAN, SRA-MSRGAN and SRA-CVEGAN), as shown in Table 3 (in the main paper).

\subsection{Additional Perceptual Comparisons}
\label{sec:comparisons}

\begin{figure*}[htbp]
\centering
\scriptsize
\centering
\begin{minipage}[b]{0.19\linewidth}
\centering
\centerline{\includegraphics[width=1.01\linewidth]{figures/partyscene_pp_orig.png}}
Original 
\end{minipage}
\begin{minipage}[b]{0.19\linewidth}
\centering
\centerline{\includegraphics[width=1.01\linewidth]{figures/partyscene_pp_orig_zoom.png}}
 Original
\end{minipage}
\begin{minipage}[b]{0.19\linewidth}
\centering
\centerline{\includegraphics[width=1.01\linewidth]{figures/partyscene_pp_anchor.png}}
 HM 16.20, QP=37
\end{minipage}
\begin{minipage}[b]{0.19\linewidth}
\centering
\centerline{\includegraphics[width=1.01\linewidth]{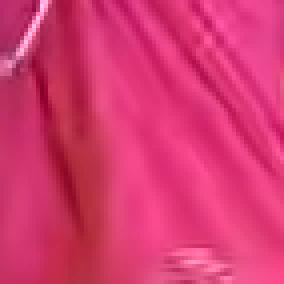}}
RCAN \cite{zhang2018image}
\end{minipage}
\begin{minipage}[b]{0.19\linewidth}
\centering
\centerline{\includegraphics[width=1.01\linewidth]{figures/partyscene_pp_rnan.png}}
RNAN \cite{zhang2019nonlocal}
\end{minipage}

\begin{minipage}[b]{0.19\linewidth}
\centering
\centerline{\includegraphics[width=1.01\linewidth]{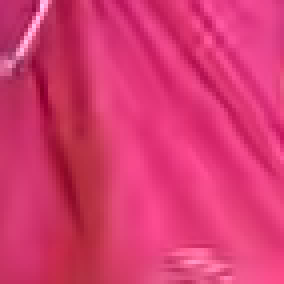}}
RCAN-GAN \cite{ren2020rcangan}
\end{minipage}
\begin{minipage}[b]{0.19\linewidth}
\centering
\centerline{\includegraphics[width=1.01\linewidth]{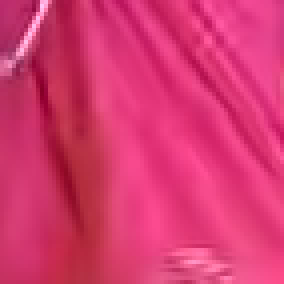}}
PatchESRGAN \cite{ji2020patchesrgan}
\end{minipage}
\begin{minipage}[b]{0.19\linewidth}
\centering
\centerline{\includegraphics[width=1.01\linewidth]{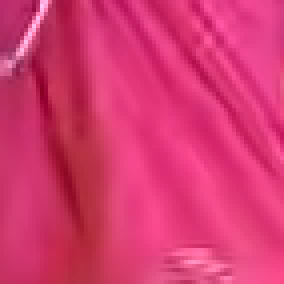}}
RFB-ESRGAN \cite{shang2020rfbesrgan}
\end{minipage}
\begin{minipage}[b]{0.19\linewidth}
\centering
\centerline{\includegraphics[width=1.01\linewidth]{figures/partyscene_pp_msrgan.png}}
MSRGAN \cite{ma2019perceptually}
\end{minipage}
\begin{minipage}[b]{0.19\linewidth}
\centering
\centerline{\includegraphics[width=1.01\linewidth]{figures/partyscene_pp_cvegan.png}}
CVEGAN (Ours)
\end{minipage}

\begin{minipage}[b]{0.19\linewidth}
\centering
\centerline{\includegraphics[width=1.01\linewidth]{figures/catrobot_orig.png}}
Original 
\end{minipage}
\begin{minipage}[b]{0.19\linewidth}
\centering
\centerline{\includegraphics[width=1.01\linewidth]{figures/catrobot_orig_zoom.png}}
Original
\end{minipage}
\begin{minipage}[b]{0.19\linewidth}
\centering
\centerline{\includegraphics[width=1.01\linewidth]{figures/catrobot_anchor.png}}
HM 16.20, QP=37
\end{minipage}
\begin{minipage}[b]{0.19\linewidth}
\centering
\centerline{\includegraphics[width=1.01\linewidth]{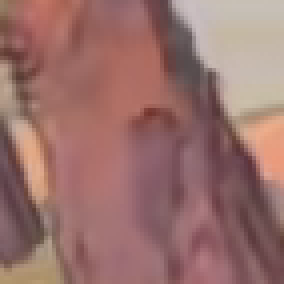}}
RCAN \cite{zhang2018image}
\end{minipage}
\begin{minipage}[b]{0.19\linewidth}
\centering
\centerline{\includegraphics[width=1.01\linewidth]{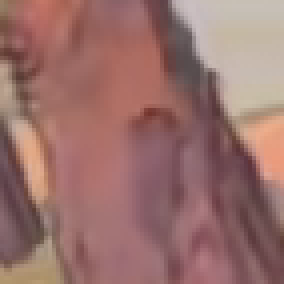}}
RNAN \cite{zhang2019nonlocal}
\end{minipage}

\begin{minipage}[b]{0.19\linewidth}
\centering
\centerline{\includegraphics[width=1.01\linewidth]{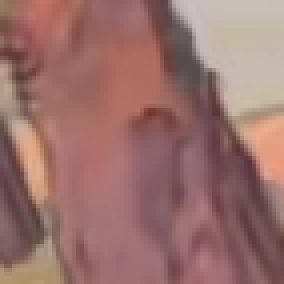}}
RCAN-GAN \cite{ren2020rcangan}
\end{minipage}
\begin{minipage}[b]{0.19\linewidth}
\centering
\centerline{\includegraphics[width=1.01\linewidth]{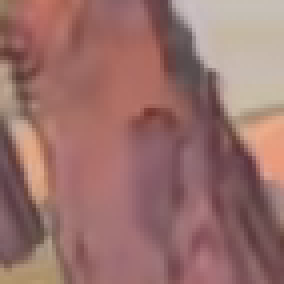}}
PatchESRGAN \cite{ji2020patchesrgan}
\end{minipage}
\begin{minipage}[b]{0.19\linewidth}
\centering
\centerline{\includegraphics[width=1.01\linewidth]{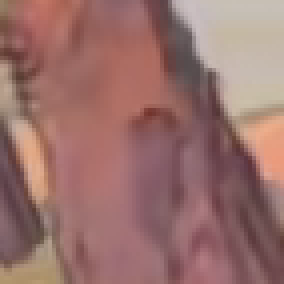}}
RFB-ESRGAN \cite{shang2020rfbesrgan}
\end{minipage}
\begin{minipage}[b]{0.19\linewidth}
\centering
\centerline{\includegraphics[width=1.01\linewidth]{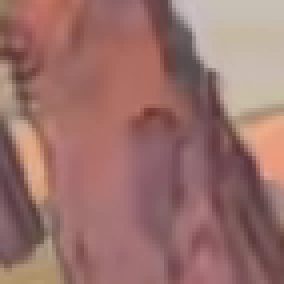}}
MSRGAN \cite{ma2019perceptually}
\end{minipage}
\begin{minipage}[b]{0.19\linewidth}
\centering
\centerline{\includegraphics[width=1.01\linewidth]{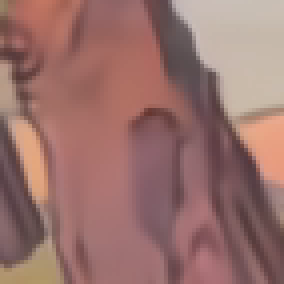}}
CVEGAN (Ours)
\end{minipage}

\caption{Two sets of example blocks cropped from the reconstructed frames generated by the anchor HM 16.20 (QP=37), six state-of-the-art network architectures and the proposed CVEGAN for CNN-based PP. The bit consumption in each example set is identical for all tested versions. Row 1 and 2 correspond to the 170th frame of the \textit{PartyScene} sequence and Row 3 and 4 correspond to the 104th frame of the \textit{CatRobot1} sequence. It can be observed that the output of CVEGAN exhibits improved perceptual quality compared to the anchor HEVC HM 16.20 and other compared networks, with fewer blocking artefacts, more textural detail and higher contrast.}

\label{fig:perceptualcom_pp1}
\end{figure*}

\begin{figure*}[htbp]
\centering
\scriptsize
\centering
\begin{minipage}[b]{0.19\linewidth}
\centering
\centerline{\includegraphics[width=1.01\linewidth]{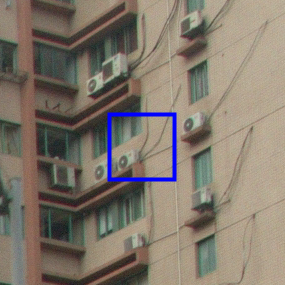}}
Original 
\end{minipage}
\begin{minipage}[b]{0.19\linewidth}
\centering
\centerline{\includegraphics[width=1.01\linewidth]{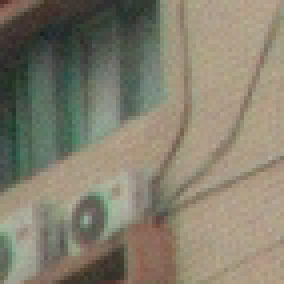}}
Original
\end{minipage}
\begin{minipage}[b]{0.19\linewidth}
\centering
\centerline{\includegraphics[width=1.01\linewidth]{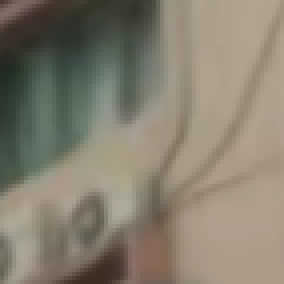}}
HM 16.20, QP=37
\end{minipage}
\begin{minipage}[b]{0.19\linewidth}
\centering
\centerline{\includegraphics[width=1.01\linewidth]{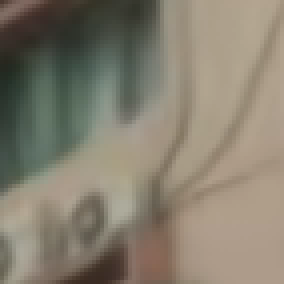}}
RCAN \cite{zhang2018image}
\end{minipage}
\begin{minipage}[b]{0.19\linewidth}
\centering
\centerline{\includegraphics[width=1.01\linewidth]{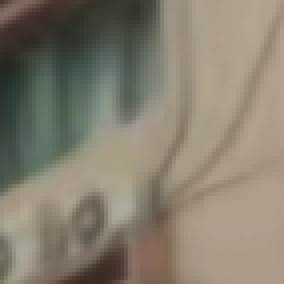}}
RNAN \cite{zhang2019nonlocal}
\end{minipage}

\begin{minipage}[b]{0.19\linewidth}
\centering
\centerline{\includegraphics[width=1.01\linewidth]{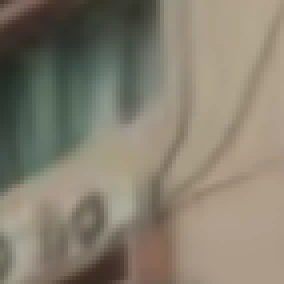}}
RCAN-GAN \cite{ren2020rcangan}
\end{minipage}
\begin{minipage}[b]{0.19\linewidth}
\centering
\centerline{\includegraphics[width=1.01\linewidth]{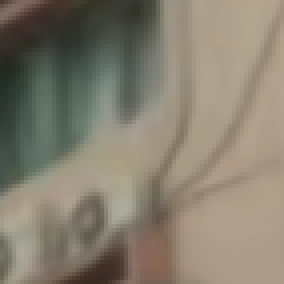}}
 PatchESRGAN \cite{ji2020patchesrgan}
\end{minipage}
\begin{minipage}[b]{0.19\linewidth}
\centering
\centerline{\includegraphics[width=1.01\linewidth]{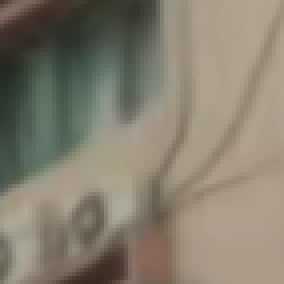}}
RFB-ESRGAN \cite{shang2020rfbesrgan}
\end{minipage}
\begin{minipage}[b]{0.19\linewidth}
\centering
\centerline{\includegraphics[width=1.01\linewidth]{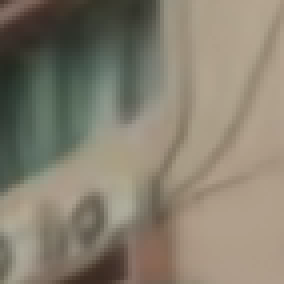}}
MSRGAN \cite{ma2019perceptually}
\end{minipage}
\begin{minipage}[b]{0.19\linewidth}
\centering
\centerline{\includegraphics[width=1.01\linewidth]{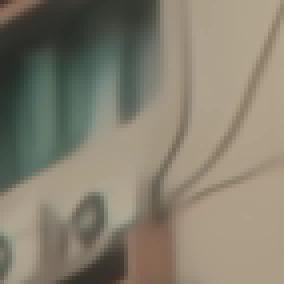}}
CVEGAN (Ours)
\end{minipage}

\begin{minipage}[b]{0.19\linewidth}
\centering
\centerline{\includegraphics[width=1.01\linewidth]{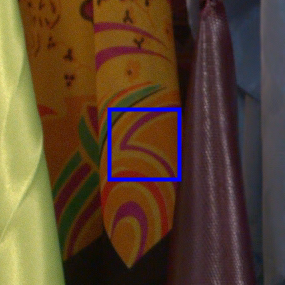}}
Original 
\end{minipage}
\begin{minipage}[b]{0.19\linewidth}
\centering
\centerline{\includegraphics[width=1.01\linewidth]{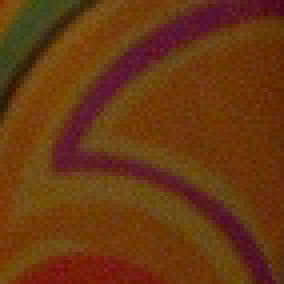}}
 Original
\end{minipage}
\begin{minipage}[b]{0.19\linewidth}
\centering
\centerline{\includegraphics[width=1.01\linewidth]{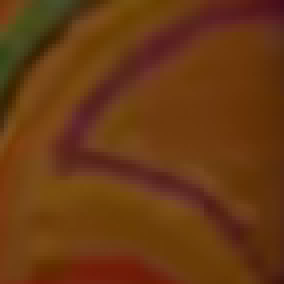}}
 HM 16.20, QP=37
\end{minipage}
\begin{minipage}[b]{0.19\linewidth}
\centering
\centerline{\includegraphics[width=1.01\linewidth]{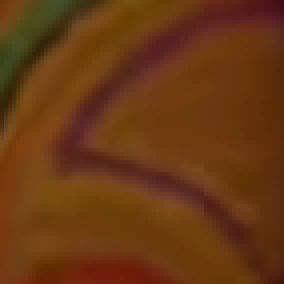}}
RCAN \cite{zhang2018image}
\end{minipage}
\begin{minipage}[b]{0.19\linewidth}
\centering
\centerline{\includegraphics[width=1.01\linewidth]{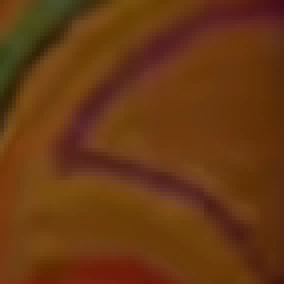}}
RNAN \cite{zhang2019nonlocal}
\end{minipage}

\begin{minipage}[b]{0.19\linewidth}
\centering
\centerline{\includegraphics[width=1.01\linewidth]{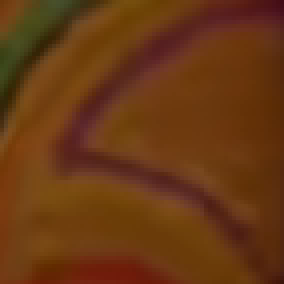}}
RCAN-GAN \cite{ren2020rcangan}
\end{minipage}
\begin{minipage}[b]{0.19\linewidth}
\centering
\centerline{\includegraphics[width=1.01\linewidth]{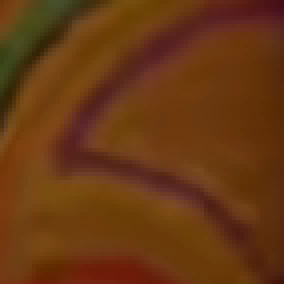}}
PatchESRGAN \cite{ji2020patchesrgan}
\end{minipage}
\begin{minipage}[b]{0.19\linewidth}
\centering
\centerline{\includegraphics[width=1.01\linewidth]{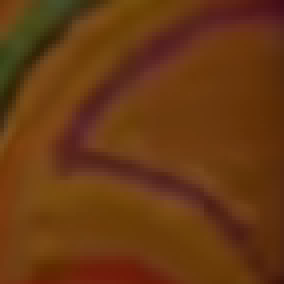}}
RFB-ESRGAN \cite{shang2020rfbesrgan}
\end{minipage}
\begin{minipage}[b]{0.19\linewidth}
\centering
\centerline{\includegraphics[width=1.01\linewidth]{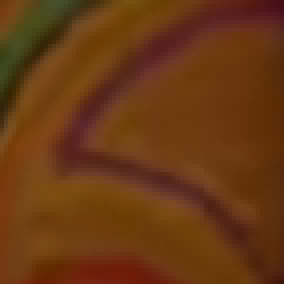}}
MSRGAN \cite{ma2019perceptually}
\end{minipage}
\begin{minipage}[b]{0.19\linewidth}
\centering
\centerline{\includegraphics[width=1.01\linewidth]{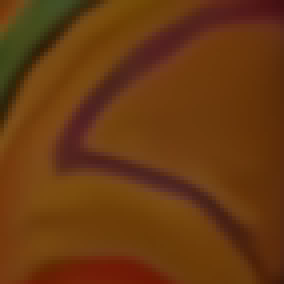}}
CVEGAN (Ours)
\end{minipage}

\caption{Two sets of example blocks cropped from the reconstructed frames generated by the anchor HM 16.20 (QP=37), six state-of-the-art network architectures and the proposed CVEGAN for CNN-based PP. The bit consumption in each example set is identical for all tested versions. Row 1 and 2 correspond to the 250th frame of the \textit{DaylightRoad2} sequence, Row 3 and 4 correspond to the 216th frame of the \textit{CatRobot1} sequence. It can be observed that the output of CVEGAN exhibits improved perceptual quality compared to the anchor HEVC HM 16.20 and other compared networks, with fewer blocking artefacts, more textural detail and higher contrast.}

\label{fig:perceptualcom_pp2}
\end{figure*}

\begin{figure*}[htbp]
\centering
\scriptsize
\centering
\begin{minipage}[b]{0.19\linewidth}
\centering
\centerline{\includegraphics[width=1.01\linewidth]{figures/catrobot_orig.png}}
Original 
\end{minipage}
\begin{minipage}[b]{0.19\linewidth}
\centering
\centerline{\includegraphics[width=1.01\linewidth]{figures/catrobot_orig_zoom.png}}
Original
\end{minipage}
\begin{minipage}[b]{0.19\linewidth}
\centering
\centerline{\includegraphics[width=1.01\linewidth]{figures/catrobot_anchor.png}}
HM 16.20, QP=37
\end{minipage}
\begin{minipage}[b]{0.19\linewidth}
\centering
\centerline{\includegraphics[width=1.01\linewidth]{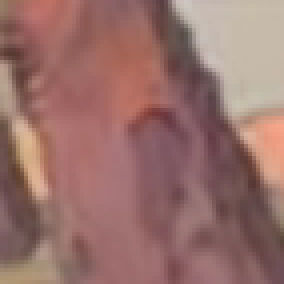}}
RCAN \cite{zhang2018image}
\end{minipage}
\begin{minipage}[b]{0.19\linewidth}
\centering
\centerline{\includegraphics[width=1.01\linewidth]{figures/catrobot_rnan.png}}
RNAN \cite{zhang2019nonlocal}
\end{minipage}

\begin{minipage}[b]{0.19\linewidth}
\centering
\centerline{\includegraphics[width=1.01\linewidth]{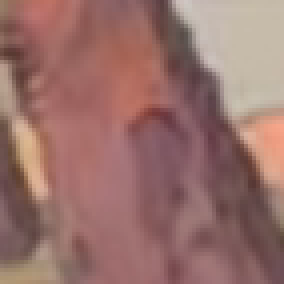}}
RCAN-GAN \cite{ren2020rcangan}
\end{minipage}
\begin{minipage}[b]{0.19\linewidth}
\centering
\centerline{\includegraphics[width=1.01\linewidth]{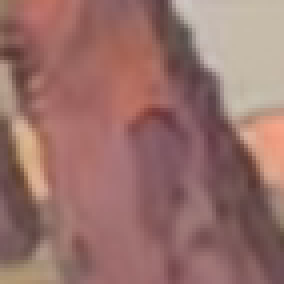}}
PatchESRGAN \cite{ji2020patchesrgan}
\end{minipage}
\begin{minipage}[b]{0.19\linewidth}
\centering
\centerline{\includegraphics[width=1.01\linewidth]{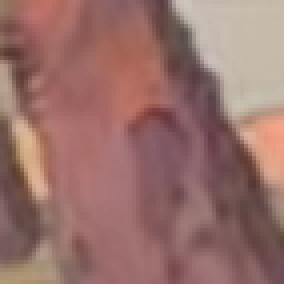}}
RFB-ESRGAN \cite{shang2020rfbesrgan}
\end{minipage}
\begin{minipage}[b]{0.19\linewidth}
\centering
\centerline{\includegraphics[width=1.01\linewidth]{figures/catrobot_msrgan.png}}
MSRGAN \cite{ma2019perceptually}
\end{minipage}
\begin{minipage}[b]{0.19\linewidth}
\centering
\centerline{\includegraphics[width=1.01\linewidth]{figures/catrobot_cvegan.png}}
CVEGAN (Ours)
\end{minipage}

\begin{minipage}[b]{0.19\linewidth}
\centering
\centerline{\includegraphics[width=1.01\linewidth]{figures/daylightroad_pp_orig.png}}
Original 
\end{minipage}
\begin{minipage}[b]{0.19\linewidth}
\centering
\centerline{\includegraphics[width=1.01\linewidth]{figures/daylightroad_pp_orig_zoom.png}}
 Original
\end{minipage}
\begin{minipage}[b]{0.19\linewidth}
\centering
\centerline{\includegraphics[width=1.01\linewidth]{figures/daylightroad_pp_anchor.png}}
HM 16.20, QP=37
\end{minipage}
\begin{minipage}[b]{0.19\linewidth}
\centering
\centerline{\includegraphics[width=1.01\linewidth]{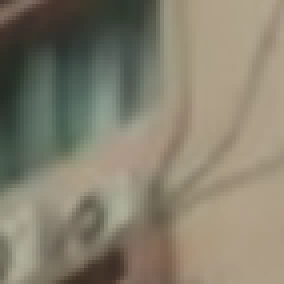}}
RCAN \cite{zhang2018image}
\end{minipage}
\begin{minipage}[b]{0.19\linewidth}
\centering
\centerline{\includegraphics[width=1.01\linewidth]{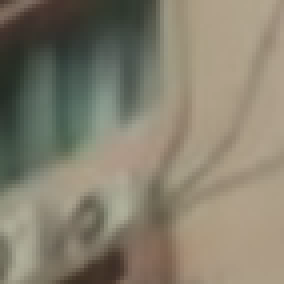}}
RNAN \cite{zhang2019nonlocal}
\end{minipage}

\begin{minipage}[b]{0.19\linewidth}
\centering
\centerline{\includegraphics[width=1.01\linewidth]{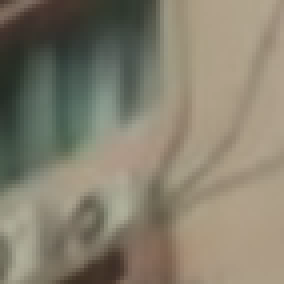}}
RCAN-GAN \cite{ren2020rcangan}
\end{minipage}
\begin{minipage}[b]{0.19\linewidth}
\centering
\centerline{\includegraphics[width=1.01\linewidth]{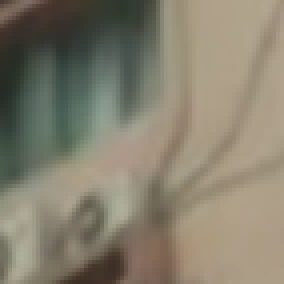}}
PatchESRGAN \cite{ji2020patchesrgan}
\end{minipage}
\begin{minipage}[b]{0.19\linewidth}
\centering
\centerline{\includegraphics[width=1.01\linewidth]{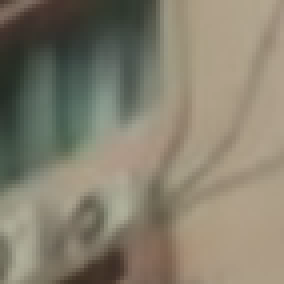}}
RFB-ESRGAN \cite{shang2020rfbesrgan}
\end{minipage}
\begin{minipage}[b]{0.19\linewidth}
\centering
\centerline{\includegraphics[width=1.01\linewidth]{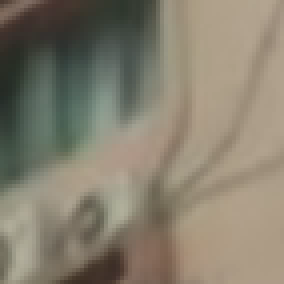}}
 MSRGAN \cite{ma2019perceptually}
\end{minipage}
\begin{minipage}[b]{0.19\linewidth}
\centering
\centerline{\includegraphics[width=1.01\linewidth]{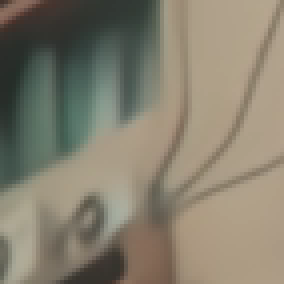}}
CVEGAN (Ours)
\end{minipage}

\caption{Two sets of example blocks cropped from the reconstructed frames generated by the anchor HM 16.20 (QP=37), six state-of-the-art network architectures and the proposed CVEGAN for CNN-based SRA. The bit consumption in each example set is similar for all tested versions. Row 1 and 2 correspond to the 104th frame of the \textit{CatRobot1} sequence, Row 3 and 4 correspond to the 250th frame of the \textit{DaylightRoad2} sequence. It can be observed that the output of CVEGAN exhibits improved perceptual quality compared to the anchor HEVC HM 16.20 and other networks, with fewer blocking artefacts, more textural detail and higher contrast.}

\label{fig:perceptualcom_sra1}
\end{figure*}

\begin{figure*}[htbp]
\centering
\scriptsize
\centering
\begin{minipage}[b]{0.19\linewidth}
\centering
\centerline{\includegraphics[width=1.01\linewidth]{figures/catrobot2_orig.png}}
Original 
\end{minipage}
\begin{minipage}[b]{0.19\linewidth}
\centering
\centerline{\includegraphics[width=1.01\linewidth]{figures/catrobot2_orig_zoom.png}}
 Original
\end{minipage}
\begin{minipage}[b]{0.19\linewidth}
\centering
\centerline{\includegraphics[width=1.01\linewidth]{figures/catrobot2_anchor.png}}
 HM 16.20, QP=37
\end{minipage}
\begin{minipage}[b]{0.19\linewidth}
\centering
\centerline{\includegraphics[width=1.01\linewidth]{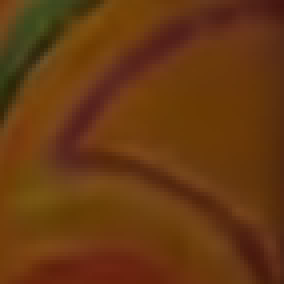}}
RCAN \cite{zhang2018image}
\end{minipage}
\begin{minipage}[b]{0.19\linewidth}
\centering
\centerline{\includegraphics[width=1.01\linewidth]{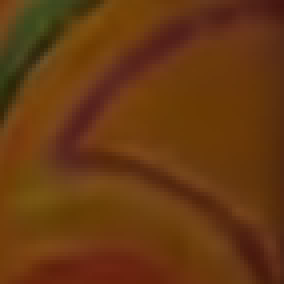}}
RNAN \cite{zhang2019nonlocal}
\end{minipage}

\begin{minipage}[b]{0.19\linewidth}
\centering
\centerline{\includegraphics[width=1.01\linewidth]{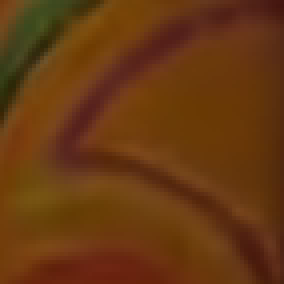}}
RCAN-GAN \cite{ren2020rcangan}
\end{minipage}
\begin{minipage}[b]{0.19\linewidth}
\centering
\centerline{\includegraphics[width=1.01\linewidth]{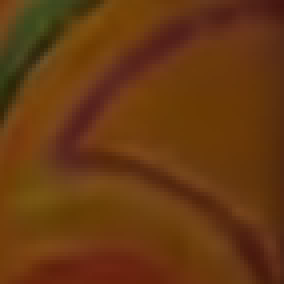}}
PatchESRGAN \cite{ji2020patchesrgan}
\end{minipage}
\begin{minipage}[b]{0.19\linewidth}
\centering
\centerline{\includegraphics[width=1.01\linewidth]{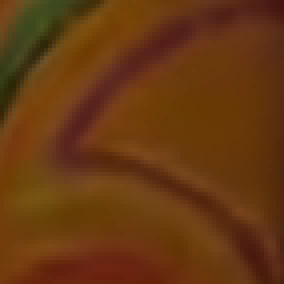}}
RFB-ESRGAN \cite{shang2020rfbesrgan}
\end{minipage}
\begin{minipage}[b]{0.19\linewidth}
\centering
\centerline{\includegraphics[width=1.01\linewidth]{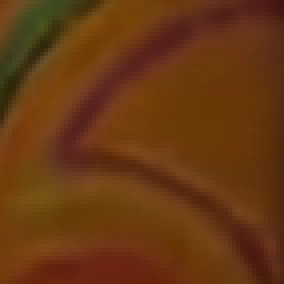}}
MSRGAN \cite{ma2019perceptually}
\end{minipage}
\begin{minipage}[b]{0.19\linewidth}
\centering
\centerline{\includegraphics[width=1.01\linewidth]{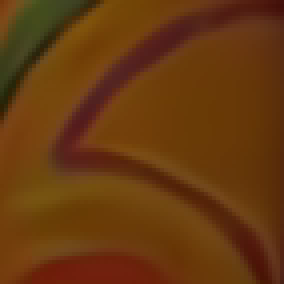}}
CVEGAN (Ours)
\end{minipage}

\begin{minipage}[b]{0.19\linewidth}
\centering
\centerline{\includegraphics[width=1.01\linewidth]{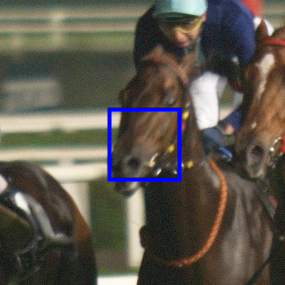}}
Original 
\end{minipage}
\begin{minipage}[b]{0.19\linewidth}
\centering
\centerline{\includegraphics[width=1.01\linewidth]{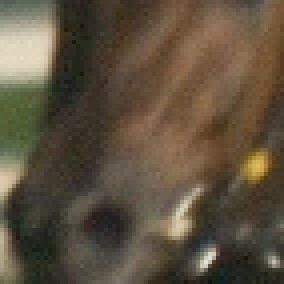}}
Original
\end{minipage}
\begin{minipage}[b]{0.19\linewidth}
\centering
\centerline{\includegraphics[width=1.01\linewidth]{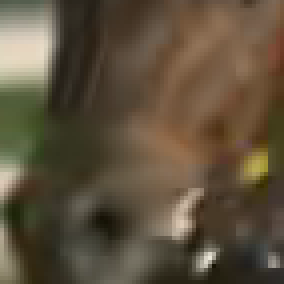}}
HM 16.20, QP=37
\end{minipage}
\begin{minipage}[b]{0.19\linewidth}
\centering
\centerline{\includegraphics[width=1.01\linewidth]{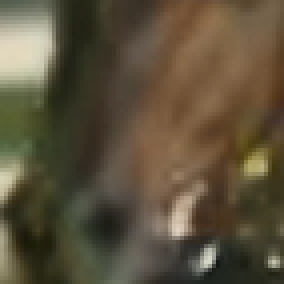}}
RCAN \cite{zhang2018image}
\end{minipage}
\begin{minipage}[b]{0.19\linewidth}
\centering
\centerline{\includegraphics[width=1.01\linewidth]{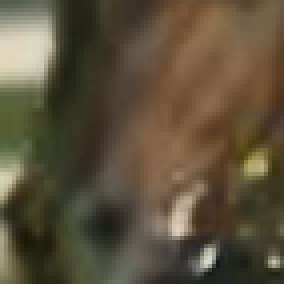}}
RNAN \cite{zhang2019nonlocal}
\end{minipage}

\begin{minipage}[b]{0.19\linewidth}
\centering
\centerline{\includegraphics[width=1.01\linewidth]{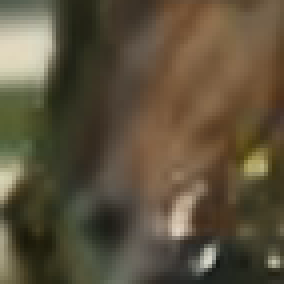}}
RCAN-GAN \cite{ren2020rcangan}
\end{minipage}
\begin{minipage}[b]{0.19\linewidth}
\centering
\centerline{\includegraphics[width=1.01\linewidth]{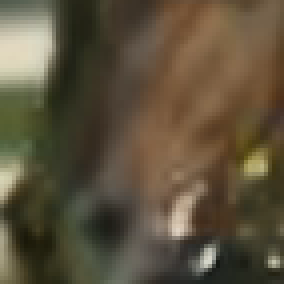}}
PatchESRGAN \cite{ji2020patchesrgan}
\end{minipage}
\begin{minipage}[b]{0.19\linewidth}
\centering
\centerline{\includegraphics[width=1.01\linewidth]{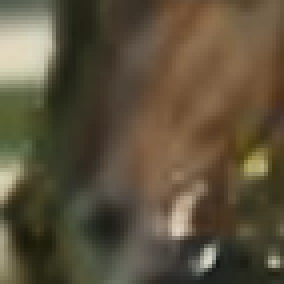}}
RFB-ESRGAN \cite{shang2020rfbesrgan}
\end{minipage}
\begin{minipage}[b]{0.19\linewidth}
\centering
\centerline{\includegraphics[width=1.01\linewidth]{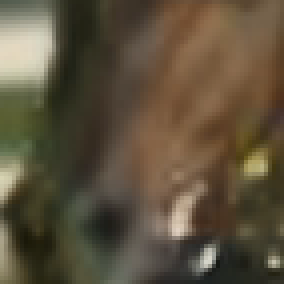}}
MSRGAN \cite{ma2019perceptually}
\end{minipage}
\begin{minipage}[b]{0.19\linewidth}
\centering
\centerline{\includegraphics[width=1.01\linewidth]{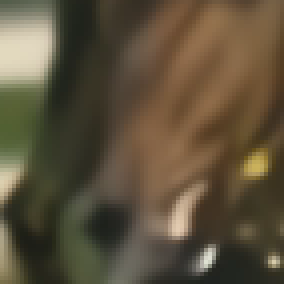}}
CVEGAN (Ours)
\end{minipage}

\caption{Two sets of example blocks cropped from the reconstructed frames generated by the anchor HM 16.20 (QP=37), six state-of-the-art network architectures and the proposed CVEGAN for CNN-based SRA. The bit consumption in each example set is similar for all tested versions. Row 1 and 2 correspond to the 216th frame of the \textit{CatRobot1} sequence and Row 3 and 4 correspond to the 161st frame of the \textit{RaceNight} sequence. It can be observed that the output of CVEGAN exhibits improved perceptual quality compared to the anchor HEVC HM 16.20 and other networks, with fewer blocking artefacts, more textural detail and higher contrast.}

\label{fig:perceptualcom_sra2}
\end{figure*}

Figures \ref{fig:perceptualcom_pp1}-\ref{fig:perceptualcom_sra2} present additional subjective comparison results among the proposed CVEGAN, the anchor HEVC and other top performing architectures for both PP and SRA. The perceptual quality improvements associated with CVEGAN can be clearly observed in these examples and provide further validation of our approach.

\end{document}